\documentclass[aps,notitlepage,nofootinbib]{revtex4-2}

\usepackage{bm,amsmath,amssymb,graphicx}
\usepackage{ amsfonts, bm, epsfig, xcolor, algorithm2e}    
\usepackage{float}
\usepackage[format=plain,labelfont={bf,small},textfont=small,justification=raggedright,singlelinecheck=false]{caption}
\usepackage{subcaption}
\usepackage{footnote}
\usepackage{tikz}

\begin{document}

\title{Some explorations of Kirchhoff dynamics}

\author{O. W. Oguns} 
\author{J. A. Hanna} \email{jhanna@unr.edu}%
\affiliation{Mechanical Engineering, University of Nevada, Reno, NV 89557-0312, U.S.A. }%

\date{\today}

\begin{abstract}
The classical problem of a rigid body in an ideal fluid is a six-dimensional dynamical system with three conserved quantities.  Despite its long history and its importance as a reduced model of many fluid-structure problems, basic questions remain, including the shapes and connectivities of the three-dimensional submanifolds of solutions, and the nature of chaotic motions on these. We explore the problem with a particular ellipsoidal body, beginning with integrable motions--- steady linear translation and rotation and periodic planar tumbling and fluttering.  The stability of these states depends on the relative magnitudes of linear and angular momentum or energy.  Linear stability, including Floquet, analysis of lower-dimensional perturbed systems is consistent with direct integration of the full system.  We document instabilities leading to a variety of regular and chaotic motions, including flipping and twirling, whose trajectories appear to follow paths near connections between the integrable states. Such indirect observation of these connections provides insight into the structure underlying the rich dynamics of this simple system.  For the submanifold where linear and angular momentum are perpendicular, we further examine how solutions fit together to fill momentum space in different zones of stability of the integrable states, and thereby uncover additional bifurcations in connectivity and what appear to be new integrable solutions. 
\end{abstract}

\maketitle

\section{Introduction}

The motion of a rigid body in an ideal fluid 
is a classical problem at the core of many engineered and natural systems. 
The Kirchhoff equations \cite{Kirchhoff1870} are an elegant reduced model describing the time evolution of the body's linear and angular momentum in the body's frame, comprising six ODEs with three conserved quantities for a generic shape: the kinetic energy and the projections of the linear momentum onto itself and onto the angular momentum (hereafter the ``angular invariant''). 
Notable work on these equations includes that of Clebsch \cite{Clebsch1871} and Lamb \cite{lambBOOK} and, about a century later, that of Kozlov \cite{Kozlov89}, Aref and co-workers \cite{ArefJones93, roenby2010atmosphere}, Holmes and co-workers \cite{Holmes98}, Borisov and Mamaev \cite{Borisov06, BorisovMamaev06}, and Kuznetsov \cite{Kuznetsov15}, by which point the significance of non-integrability was understood \cite{Perelomov81, KozlovOniscenko82}. Some of these later works include gravity, which modifies certain aspects of integrability; the problem of a falling body was studied qualitatively by Maxwell \cite{Maxwell1854} before Kirchhoff, and later by Chaplygin. 
Additional interesting features arise for bodies with nonuniform density resulting in noncoincident centers of mass and buoyancy \cite{Mahadevan96, WeissmannPinkall12}. 
For the basic problem, the adjustable parameters are the ratio of mass density between solid and fluid, and body tensors encoding shape properties involving distribution of mass and surface area, which produce linear and angular inertias intrinsic to the body and inherited from interaction with the fluid. 
Qualitatively, the latter introduce anisotropy into the linear momentum and, for some bodies, cross-coupling between linear and angular velocities and momenta \cite{WeissmannPinkall12}. 
Too many other works to mention have explored special cases in which a nongenerically symmetric body introduces an additional conserved quantity and makes the system integrable. 
Further interesting reviews and commentary on the Kirchhoff problem may be found in the books \cite{BorisovMamaevBOOK, ShashikanthBOOK}. 

Image sequences from experiments on falling or rising objects, including chaotic fluttering and tumbling, are shown in various works 
 \cite{Schmiedel28, Dupleich41, Willmarth64, Stringham69, Field97, Belmonte98, Mahadevan99, Andersen05-1, Fernandes05, Zhong11, Wang13, Varshney13,  Heisinger14}. 
Most aspects of these phenomena are already present in the unadorned Kirchhoff equations, with or even without gravity. 
 However, some experiments on thin bodies indicate deviations in phase between body orientation and trajectory that are not captured by the Kirchhoff equations \cite{Fernandes05, Andersen05-1}. 
 And it has been stated \cite{Mahadevan99} that the dependence of observed tumbling frequencies on aspect ratio is inconsistent with a purely inviscid theory. 
There have been numerous studies of reduced models, of higher complexity than Kirchhoff equations, for thin plates falling under gravity, which we do not cover here.

The present work is inspired by the largely unexplored space of solutions of the Kirchhoff problem. 
It is surprising how little this skeleton of rigid-body-fluid dynamics has been mapped out in detail. 
A few integrable motions are known for the Kirchhoff equations, and serve as preliminary landmarks for our study.  The stability of steady translations and rotations about the same fixed axis of symmetry (``pure'') was determined by Holmes and co-workers \cite{Holmes98}, alongside more complex steady motions that exist for certain classes of bodies. 
Two types of time-dependent angular momentum around a constant axis perpendicular to the linear momentum, which we will call ``planar'' tumbling and fluttering, were known to Lamb \cite[Article 127]{lambBOOK}, and studied by Aref and Roenby \cite{roenby2010atmosphere} and Kuznetsov \cite{Kuznetsov15}, but only as strictly two-dimensional dynamics--- the stability of these motions as three-dimensional solutions was not considered, although Lamb \cite[Article 128]{lambBOOK} examined the stability of nonrotating ellipsoids and rotating solids of revolution. 

Little is known about transitions from integrable to chaotic behavior in this system. 
Chaos was first numerically demonstrated by Aref and Jones \cite{ArefJones93}, who restricted themselves to the case of vanishing angular invariant. 
For both the Kirchhoff equations and full numerical simulations of the surrounding fluid, Essmann and co-workers \cite{essmann2020chaotic} examined the propensity for chaos in terms of density ratio and initial conditions reflecting the translational to rotational kinetic energy ratio. 
Borisov and Mamaev \cite{BorisovMamaev06} examined transition in a related planar problem. 
Apparently chaotic behavior near the border between tumbling and fluttering has been observed in experiments with gravity \cite{Stringham69, Field97, Belmonte98, Andersen05-1}. 

The six ODEs and three conserved quantities define a three-dimensional space in which the solutions live.  The shapes and connectivities of the possible spaces, and how these intersections change with the conserved quantities, have to our knowledge not been described. Topologies and bifurcation sets have been determined for lower-dimensional cases, either in the presence of additional symmetries \cite{OrelRyabov98, dragovic2022topology}, or for planar dynamics with added gravity and circulation terms \cite{BorisovMamaev06}. 

In the present work, we begin to explore the space of solutions of the Kirchhoff equations. We employ a simple uniform shape, an ellipsoid with three distinct axes, originally introduced by Aref and Jones \cite{ArefJones93} and recently employed by Essmann and co-workers \cite{essmann2020chaotic}.  We set the mass density ratio to unity, because the latter work indicates that we should thereby expect to obtain a somewhat even mixture of quasiperiodic and chaotic orbits.  The equations and body are introduced in Section \ref{equations}. In Section \ref{landmark}, we determine the stability of known integrable landmark solutions, and in Section \ref{examplesolutions} we note the nature of these instabilities and present various examples of resulting trajectories. We observe a transition to chaos involving irregular sequences of paths between unstable planar landmarks. Regular or chaotic trajectories along these paths indirectly reveal connectivity of the solution space. We also see more complex chaotic motions, like those already observed. In Section \ref{fillspace}, we consider vanishing angular invariant, and examine how various types of solutions fit together to fill spaces in different zones corresponding to the presence and stability of the planar modes. This further reveals additional connectivity bifurcations and types of integrable motion.  Discussion follows in Section \ref{discussion}. 

Our exploration of the solution space, even for this one choice of shape and mass ratio, remains incomplete as we explore only values of the angular invariant close to maximal (when we perturb near the pure modes) or close to zero (when we perturb near the planar modes), or strictly zero (when we fill space with trajectories).  Note that planar modes are only a special subcase of zero angular invariant solutions; the entire set of such solutions is much larger than this, including many fully three-dimensional regular and chaotic orbits, including those shown by Aref and Jones \cite{ArefJones93} and in the present work. We also concern ourselves primarily with trajectories in momentum space, rather than real space.

\section{Equations}\label{equations}

We consider the Kirchhoff equations in their most basic form, for a body with coincident centers of mass and buoyancy, in the absence of gravity or any other external forces and torques.  In the rotating body frame, linear and angular momentum $\bm{P}$ and $\bm{L}$ evolve according to the six equations
\begin{equation}\begin{aligned}
    \dot{\bm{P}} &= \bm{P} \times \bm{\mathit{\Omega}}\, ,  \\
    \dot{\bm{L}} &= \bm{L} \times \bm{\mathit{\Omega}} + \bm{P}\times\bm{U}\, , \label{kirch}
\end{aligned}\end{equation}
and are further related to the linear and angular velocity $\bm{U}$ and $\bm{\mathit{\Omega}}$ by  
\begin{equation}\begin{aligned}
    \bm{P} &= {\bf M} \cdot \bm{U} + {\bf S}\cdot\bm{\mathit{\Omega}}\, ,\\
    \bm{L} &= {\bf I} \cdot \bm{\mathit{\Omega}} +{\bf S}^\top\cdot\bm{U}\, ,
\end{aligned}\end{equation}
where the constant tensors $\bf{M}$, $\bf{I}$, $\bf{S}$ depend on the shape of the body, and encode both its intrinsic inertial properties as well as those inherited from interaction with the fluid; for discussion see \cite{ArefJones93, WeissmannPinkall12, CavigliaMorro17}.  
The propeller tensor $\bf{S}$ will be zero for the example body used in the present work. 
Equations \eqref{kirch} conserve three quantities: the energy $E=\tfrac{1}{2}\bm{U} \cdot \bm{P} + \tfrac{1}{2} \bm{\mathit{\Omega}} \cdot \bm{L}$ and the projections of the linear and angular momentum onto the linear momentum $\bm{P} \cdot \bm{P}$ and $\bm{L} \cdot \bm{P}$.  The former is clearly seen to be the magnitude of the linear momentum, while the latter angular invariant does not have a single standard name. 
On the three-dimensional submanifold $\bm{P} \cdot \bm{P} = 0$, the magnitude of the angular momentum $\bm{L} \cdot \bm{L}$ is conserved, and the resulting integrable system is identical in form to a purely rotating rigid body, with trajectories in angular momentum space being intersections of energy ellipsoids and angular momentum spheres. 

\subsection{The Aref-Jones ellipsoid and its solution manifold}\label{arefjones}

We employ a uniform ellipsoid with semiaxes $a=1$, $b=0.8$, $c=0.6$, and unit solid and fluid densities, $\rho_s=1$ and $\rho_f=1$.  Equations will retain dimensions. 
The total mass of the ellipsoid, which will be used to normalize some quantities later, is $M_s =\rho_s \frac{4\pi}{3}abc = 2.0106$ (here and in what follows, we write values to four decimal places, although they are calculated and used in MATLAB \cite{MATLAB:2023} to higher precision). 
The shape is simple in several ways.  Its centers of mass and buoyancy coincide, it has fore-aft symmetry, and is not chiral.  Its propeller tensor ${\bf S}$ vanishes, resulting in ``constitutive'' relations of the form 
\begin{equation}\begin{aligned}
    \bm{P} &= {\bf M} \cdot \bm{U} \, , \\
    \bm{L} &= {\bf I} \cdot \bm{\mathit{\Omega}} \, . \label{kirchAJconst}
\end{aligned}\end{equation}
Using mutual principal axes, the remaining tensors are diagonalized with elements 
$(M_a, M_b, M_c) = (2.6455, 2.9464, 3.6059)$ and $(I_a, I_b, I_c) = (0.4275, 0.6383, 0.6769)$. The eigenvalues of the two tensors have the same order, which is typical but not always so for general ellipsoids \cite{Holmes98}. 
These numbers are obtained, as presumably also in \cite{essmann2020chaotic}, using standard expressions for the mass and moment of inertia of an ellipsoid, and expressions in Lamb  \cite[Articles 114, 115, 118]{lambBOOK} for computing the fluid contributions:  
\begin{equation}\begin{aligned}
    \mathbf{M}=& 
    \begin{bmatrix}
    M_a & 0 & 0 \\ 
    0 & M_b & 0 \\ 
    0  & 0 & M_c \end{bmatrix}
    =\rho_s \frac{4\pi}{3}abc \begin{bmatrix}
    1 & 0 & 0 \\ 
    0 & 1 & 0 \\ 
    0  & 0 & 1  \end{bmatrix}
    +\rho_f \frac{4\pi}{3}abc \begin{bmatrix}
    \frac{\alpha}{2-\alpha} & 0 & 0 \\ 
    0 & \frac{\beta}{2-\beta}  & 0 \\
    0 & 0 & \frac{\gamma}{2-\gamma}  \end{bmatrix} \, ,
\\
   \mathbf{I}=& 
    \begin{bmatrix}
    I_a & 0 & 0 \\ 
    0 & I_b & 0 \\ 
    0  & 0 & I_c \end{bmatrix}
   =\rho_s\frac{4\pi}{15}abc \begin{bmatrix}
    (b^2+c^2) & 0 & 0 \\ 
    0 & (c^2+a^2) & 0 \\ 
    0  & 0 & (a^2+b^2)  \end{bmatrix} \\
   & \quad\quad\quad\quad\quad\;\;+\rho_f\frac{4\pi}{15}abc \begin{bmatrix}
     \frac{(b^2 -c ^2)^2(\gamma - \beta)}{2(b^2 - c^2)+(b^2+c^2)(\beta - \gamma)} & 0 & 0 \\  
     0 & \frac{(c ^2 - a^2 )^2(\alpha-\gamma)}{2(c^2-a^2)+(c^2+a^2)(\gamma-\alpha)}  & 0 \\  
     0  & 0 & \frac{(a^2 -b^2)^2(\beta - \alpha)}{2(a^2-b^2)+(a^2 + b^2)(\alpha - \beta)} \end{bmatrix} \, ,
\\
&\mathrm{where} \quad	\alpha \equiv  abc\int_{0}^{\infty}\frac{\mathrm{d}\lambda}{(a^2+\lambda)\Delta} \quad , \quad 
	 \beta \equiv  abc\int_{0}^{\infty}\frac{\mathrm{d}\lambda}{(b^2+\lambda)\Delta} \quad , \quad 
	 \gamma \equiv  abc\int_{0}^{\infty}\frac{\mathrm{d}\lambda}{(c^2+\lambda)\Delta} \quad , \\ 
& \quad\quad\quad\;\, \Delta \equiv \sqrt{(a^2+\lambda)(b^2+\lambda)(c^2+\lambda)} \quad .
\end{aligned}\end{equation}

For future reference, the component form of the six simplified equations is now
\begin{equation}\begin{aligned}
\dot{P}_a &= \frac{1}{I_c} P_b L_c - \frac{1}{I_b} P_c L_b \, , \\
\dot{P}_b &= \frac{1}{I_a} P_c L_a - \frac{1}{I_c} P_a L_c \, , \\
\dot{P}_c &= \frac{1}{I_b} P_a L_b - \frac{1}{I_a} P_b L_a \, , \\
\dot{L}_a &= \left( \frac{1}{I_c} - \frac{1}{I_b} \right) L_b L_c + \left( \frac{1}{M_c} - \frac{1}{M_b} \right) P_b P_c \, , \\
\dot{L}_b &= \left( \frac{1}{I_a} - \frac{1}{I_c} \right) L_c L_a + \left( \frac{1}{M_a} - \frac{1}{M_c} \right) P_c P_a \, , \\
\dot{L}_c &= \left( \frac{1}{I_b} - \frac{1}{I_a} \right) L_a L_b + \left( \frac{1}{M_b} - \frac{1}{M_a} \right) P_a P_b \, . \label{kirchcomp}
\end{aligned}\end{equation}
For this paper, these equations were numerically integrated with MATLAB's ode45 function \cite{MATLAB:2023}, with absolute and relative tolerances of $10^{-10}$, such that the conserved quantities were maintained to the sixth decimal place.

The three conserved quantities define level sets, nested five-dimensional surfaces in the space spanned by the six components of the momenta. 
An energy $E$ of the simple form $\frac{1}{2}\left( \bm{U}\cdot {\bf M} \cdot \bm{U}+ \bm{\mathit{\Omega}} \cdot {\bf I} \cdot \bm{\mathit{\Omega}} \right) = \frac{1}{2}\left( \bm{P}\cdot {\bf M}^{-1} \cdot \bm{P}+ \bm{L} \cdot {\bf I}^{-1} \cdot \bm{L} \right)$ is an ellipsoid aligned with the principal directions, with axes that depend on the eigenvalues. The linear invariant $\bm{P} \cdot \bm{P}$ has two spherical dimensions in the $P$-subspace and three cylindrical ones constituting the $L$-subspace ($S^2 \times \mathbb{R}^3$).  
The angular invariant $\bm{L} \cdot \bm{P}$ is apparently some type of hyperboloid with axes of symmetry symmetrically oblique to the principal directions, or equivalently with asymptotes aligned with the principal directions. The intersections of these surfaces are three-dimensional spaces which are filled by the possible solutions of the Kirchhoff equations. The topology of these submanifolds, and the associated bifurcations as the quantities are varied, are to our knowledge unknown.  

\section{Integrable landmark solutions}\label{landmark}

\subsection{Pure modes and their linear stability}\label{puremodestability}

The contents of this section were derived by Holmes and co-workers in \cite{Holmes98}, who coined the term ``pure modes'' for the simplest possible motions of a rigid body, and determined their stability.  Their paper also provides further detail, and many other interesting analytical and numerical results. The pure mode stability results are repeated here for completeness. 
Orbits near unstable pure modes will be shown in Section \ref{perturbpure}. 

Pure modes consist of steady translation and rotation around the same principal axis of the body.  Accordingly, the linear and angular momentum are also aligned. These solutions are fixed-point equilibria in momentum space. Given $M_c > M_b > M_a$, ellipsoids can have other orderings instead of the standard $I_c > I_b > I_a$. 
According to \cite{Holmes98}, for some range of momentum ratio, these bodies can exhibit ``mixed modes'' corresponding to translation and rotation along different axes in a plane spanned by two principal axes. 
The Aref-Jones ellipsoid has the standard ordering, so does not admit mixed mode equilibria. 

\begin{figure}[!h]
    \centering
    \includegraphics[width=0.65\textwidth]{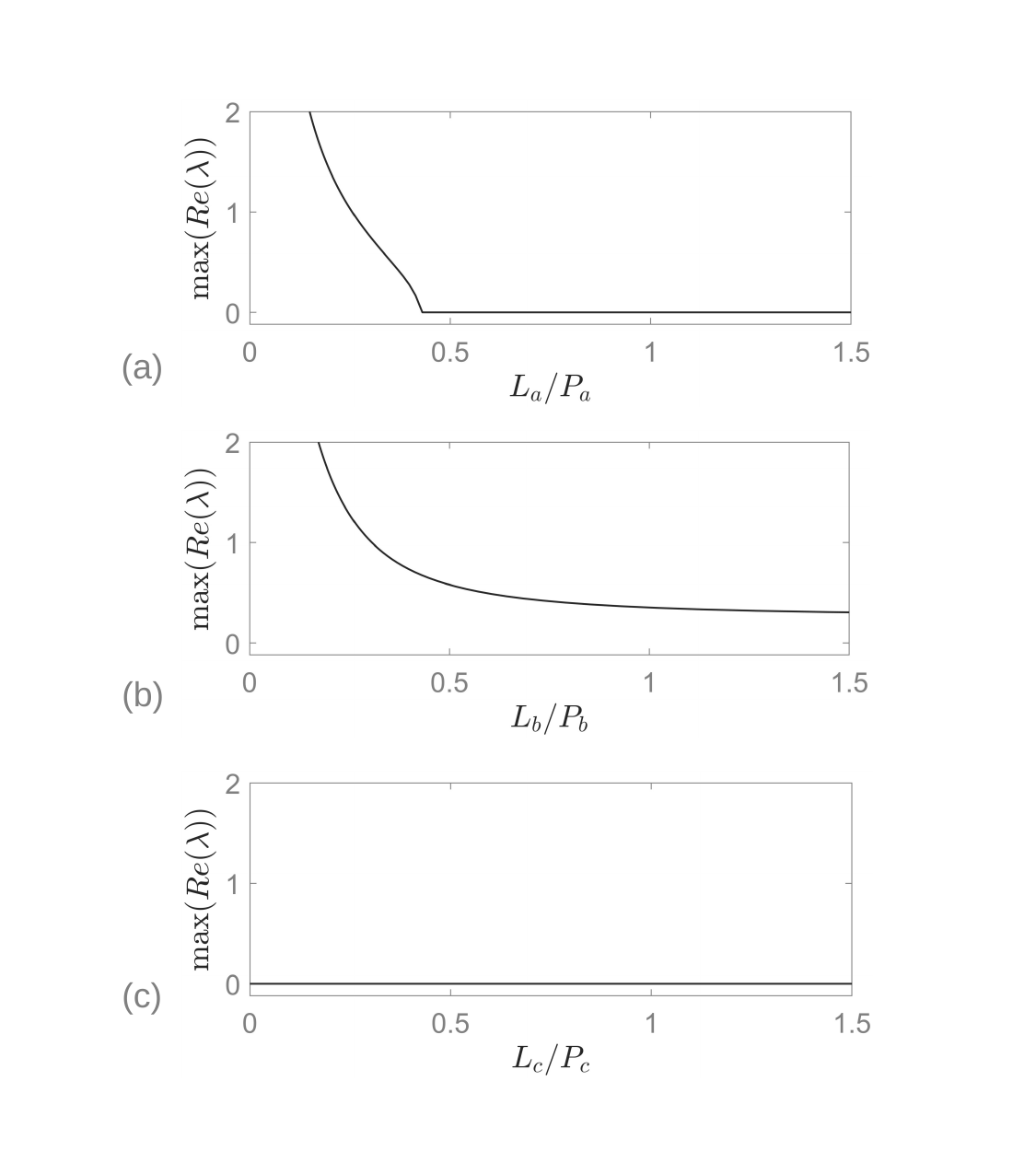}
    \vspace{-1cm}
    \caption{The largest real part of any eigenvalue for the perturbed dynamics around pure modes (a), (b), (c), in terms of relevant ratios of angular to linear momentum. Results from \cite{Holmes98}. Nonzero values indicate instability.   Note that the horizontal axes do not correspond to invariant quantities, so vertical correspondence between the plots is not meaningful.}
    \label{PmodeB}
\end{figure}

Let the components consist of a base state and a perturbed field, denoted $P_a = P_{a0} + p_a$, and so on. 
The pure mode base state consists of two constant nonzero terms. 
For example, in the pure $a$-axis mode the only nonvanishing terms in the base state are two constants $P_{a0}$ and $L_{a0}$. 
Linearizing \eqref{kirchcomp} about this solution results in a four-dimensional system, 
\begin{equation}
\begin{bmatrix}
    \dot p_b \\
    \dot p_c \\
    \dot l_b \\
    \dot l_c
\end{bmatrix}=
\begin{bmatrix}
    0 &\frac{1}{I_a}L_{a0} & 0 & -\frac{1}{I_c}P_{a0} \\
    -\frac{1}{I_a}L_{a0} & 0 & \frac{1}{I_b}P_{a0}& 0\\
0 & (\frac{1}{M_a}-\frac{1}{M_c})P_{a0} & 0 & (\frac{1}{I_a}-\frac{1}{I_c})L_{a0} \\
(\frac{1}{M_b}-\frac{1}{M_a})P_{a0} & 0 & (\frac{1}{I_b}-\frac{1}{I_a})L_{a0} & 0
\end{bmatrix}
\begin{bmatrix}
    p_b  \\
    p_c  \\
    l_b \\
    l_c
\end{bmatrix} \, , \label{pureperturb}
\end{equation}
whose eigenvalues $\lambda$ are determined by a quadratic in $\lambda^2$, yielding 
\begin{equation}\begin{aligned}
&    \lambda=\pm \sqrt{-c_2 \pm \sqrt{c_2^2-c_0}} \, , \\
&\mathrm{with} \quad    c_0=-\frac{1}{I_bI_c} \left(  \frac{I_a-I_b}{I_a^2} L_{a0}^2+\frac{M_a-M_b}{M_aM_b} P_{a0}^2  \right)  \left( \frac{I_c-I_a}{I_a^2} L_{a0}^2+\frac{M_c-M_a}{M_cM_a} P_{a0}^2  \right) \, , \\
    &\quad\quad\;\, 2c_2=\frac{1}{I_a^2}\left[ 1- \left(\frac{I_a}{I_b}-1\right)\left(1-\frac{I_a}{I_c}\right) \right]  L_{a0}^2 +\left( \frac{M_a-M_b}{M_aM_bI_c}-\frac{M_c-M_a}{M_cM_aI_b} \right) P_{a0}^2  \, .
\end{aligned}\end{equation}
When the ratio of angular to linear momentum (this has units of inverse length, but could be normalized by multiplying by the long axis $a=1$ of the object) is below a critical value,
\begin{equation}\begin{aligned}
    &\left(\frac{L_{a0}}{P_{a0}}\right)^2 < \frac{I_bI_c}{I_b+I_c-I_a} \left( 1+ \sqrt{1-\left( \frac{f_2}{f_1} \right)^2} \right) f_1 \, , \\
     &\mathrm{where} \quad f_1 \equiv -\left(\frac{1}{M_b}-\frac{1}{M_a}\right)\left(2-\frac{I_a}{I_c}\right) - \left(\frac{1}{M_a}-\frac{1}{M_c}\right)\left(\frac{I_a}{I_b}-2\right)  \, ,  \\
     & \quad\quad\quad\;\, f_2 \equiv -\left(\frac{1}{M_b}-\frac{1}{M_a}\right) \frac{I_a}{I_c} - \left(\frac{1}{M_a}-\frac{1}{M_c}\right)\frac{I_a}{I_b} \, , 
\end{aligned}\end{equation}
the system is unstable,
and when the ratio is larger, the system is (marginally) stable. 
Systems of equations corresponding to the other axes are obtained by cyclic permutation.  The pure b-axis mode is always unstable and the pure c-axis mode is always (marginally) stable. 
Figure \ref{PmodeB} shows the largest real part of the eigenvalues.  With low angular momentum, the body can execute a pure mode only along its small axis.  With sufficient rotation, the long axis pure mode is stabilized. 

\subsection{Planar modes: Tumble and Flutter}\label{planarmodes}

\begin{figure}[h]
    \centering
    \includegraphics[width=\textwidth]{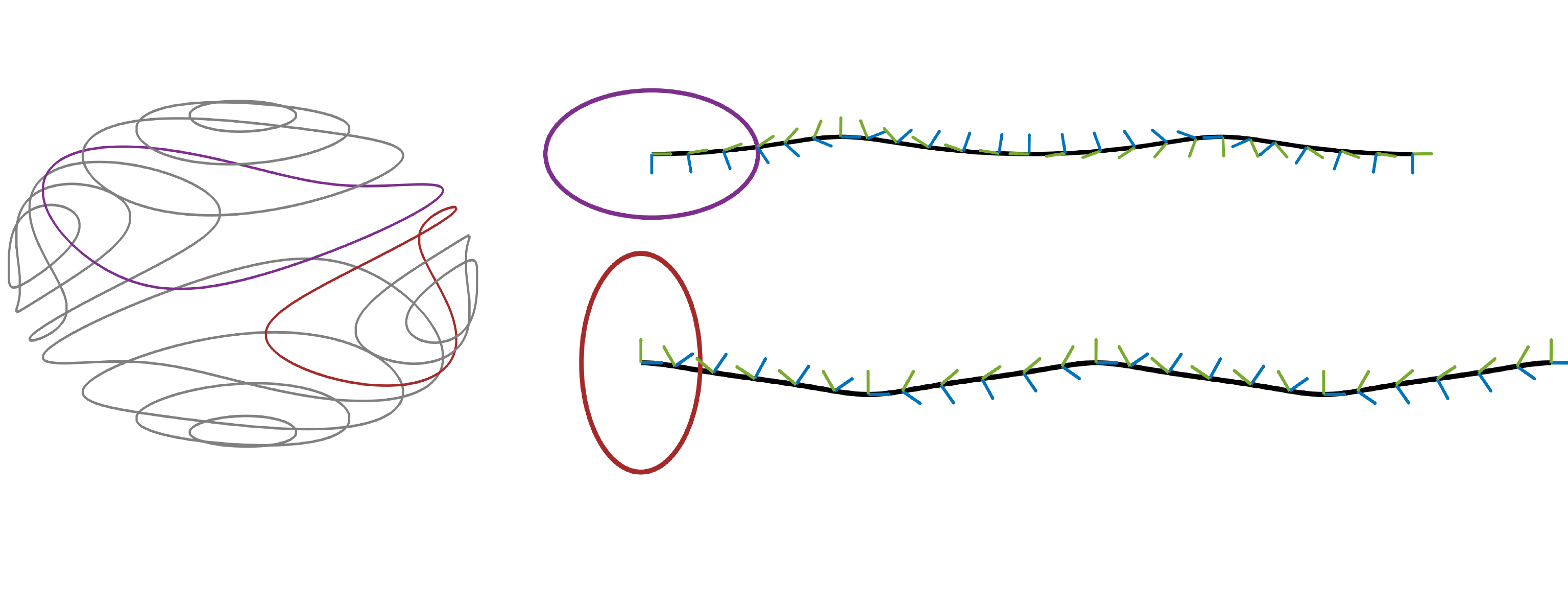}
    \vspace{-1.5cm}
    \caption{Examples of planar modes.  Left: In the space spanned by $L_b$ (oriented vertically), $P_a$, and $P_c$, several trajectories are shown, formed by the intersection of an energy ellipsoid ($E=90$) and several linear momentum cylinders.  Two types of motion are exemplified by the purple tumbling $\text{T}_b$ ($P^2=15$, $k \approx 0.822$) and maroon fluttering $\text{F}_{b,c}$ ($P^2=16$, $k \approx 1.36$) trajectories.  Flutter always aligns the smaller body axis, here $c$, with the direction of motion on average.  Right: In real space, center of mass position and $a$-$c$ frame orientations are shown for these two trajectories, along with an ellipsoid each for scale. 
 }
    \label{planarsolutions}
\end{figure}

These are motions in which the angular momentum $\bm{L}$ has a constant direction, while the linear momentum $\bm{P}$ lives in the perpendicular plane. This is a restricted integrable subcase of $\bm{L} \cdot \bm{P} = 0$, which is not integrable in general. 
The trajectories live in a three-dimensional submanifold consisting of two components of linear and one of angular momentum, and are intersections between an energy ellipsoid and a linear momentum cylinder. 
From this picture, it can be immediately seen that there are two qualitative types of generic solutions. 
Tumbling is end-over-end rotation of the body, where the orientation $\theta$ is a monotonic function of time. 
Fluttering is oscillation of $\theta$ between two limiting angles, always such that the body moves with broadside orientation, its direction of motion on average along the smaller of the two axes in the plane. The other axis corresponds to a saddle. 
We adopt a shorthand for planar states, such that $\text{T}_b$ denotes tumbling about the $b$ axis, and $\text{F}_{b,c}$ denotes flutter about the $b$ axis oscillating around translation along the $c$ axis. The latter is a redundant reminder, as there is only one option for flutter about each axis, with the three possible types being $\text{F}_{b,c}$, $\text{F}_{c,b}$, and $\text{F}_{a,c}$.  Note that these are not cyclic permutations. 
Figure \ref{planarsolutions} shows examples of iso-energy trajectories in momentum space, along with a tumble $\text{T}_b$ and a flutter $\text{F}_{b,c}$ trajectory in real space, the latter using results that follow below. 
Experimental and numerical observations of the Kirchhoff system indicate that these qualitative features persist with the addition of gravity, with the overall direction of fluttering trajectories aligned with gravity, and tumbling oblique.  

From this qualitative picture, it can also be inferred that ``in plane'' perturbations, that is, perturbations remaining on this integrable submanifold, are not relevant to the stability of solutions. In order to explore ``out of plane'' perturbations into the remaining three dimensions in the following Section \ref{planarmodestability}, we need analytical expressions for the solutions. 
Much of the below appears in some form in Kirchhoff \cite{Kirchhoff1870}, Lamb \cite[Article 127]{lambBOOK}, and Kuznetsov \cite{Kuznetsov15}. 

Let us take the example of rotation around the $b$ axis. 
Equations \eqref{kirchcomp} reduce to
\begin{equation}
    \dot P_a = -\frac{1}{I_b}P_cL_b \quad , \quad
    \dot P_c = \frac{1}{I_b}P_aL_b \quad , \quad
    \dot L_b = \left( \frac{1}{M_a}-\frac{1}{M_c}\right) P_cP_a \label{kirchplanar} \quad ,
\end{equation}
with two nontrivial conserved quantities, $P^2 = P_a^2+P_c^2$ and $E = \frac{1}{2}\left( \frac{P_a^2}{M_a}+\frac{P_c^2}{M_c}+\frac{L_b^2}{I_b} \right)$.  
These admit simplification in terms of an angle $\theta(t)$, such that $P_a= -P\sin\theta$, $P_c= P\cos\theta$, and $L_b=I_b\dot\theta$, resulting in a single pendulum-type equation,  
\begin{equation}
    \ddot \theta+ \left( \frac{M_c-M_a}{M_cM_aI_b}\right)P^2 \sin\theta \cos\theta =0 \, , \label{thetapendulum}
\end{equation}
with $2\theta$-periodicity due to the body 
symmetry. 
This immediately suggests an ansatz in terms of elliptic functions, in which $\theta$ is an amplitude $\text{am}$, its derivative and therefore $L_b$ is a $\text{dn}$ function, and $P_a$ and $P_c$ are $\text{sn}$ and $\text{cn}$ functions.  Using conservation of energy, a solution may be written in the form
\begin{equation}\begin{aligned}
	\theta &= \text{am}(\eta t, k) \quad , \quad P_a = -P\text{sn}(\eta t, k)  \quad , \quad P_c= P\text{cn}(\eta t, k)  \quad , \quad 	L_b= I_b\eta\text{dn}(\eta t, k)  \quad ,  \\
	&\mathrm{with} \quad \eta \equiv \sqrt{\frac{2E(1-\gamma)}{I_b}} 
	\quad , \quad k \equiv \sqrt{\frac{\gamma}{1-\gamma}\frac{M_c - M_a}{M_a}} \quad , \quad \gamma \equiv \frac{P^2}{2EM_c}  \quad . \label{solb}
\end{aligned}\end{equation}
The two types of orbit correspond to whether the modulus $k$ is greater or smaller than one, or equivalently when the ratio of linear momentum to energy $\frac{P^2}{2E}$ is greater or smaller than the larger coefficient of the mass tensor in the plane of motion $M_a$.   Tumbling, akin to pendulum rotation, has relatively more rotational energy and a modulus $k<1$. Fluttering, akin to pendulum libration, has relatively more translational energy and a modulus $k>1$. 
The separatrix $k=1$ has the Gudermannian function solution $\theta = \text{am}(\eta t, 1) = \text{gd}(\eta t) = 2\text{arctan}\left(\text{tanh}\frac{\eta t}{2}\right)$, 
with the further properties that 
$e^{\eta t}=\sec\theta+\tan\theta$ \cite[section 6.12.1]{CRCtables} and $\dot \theta = \eta\cos{\theta}$, the latter a form appearing in Lamb. 
The functions $\text{sn}$, $\text{cn}$, $\text{dn}$ become respectively $\text{tanh}$, $\text{sech}$, $\text{sech}$. It is also possible to rewrite these equations using certain identities that exchange the roles of $\text{cn}$ and $\text{dn}$ upon inversion of the modulus, namely $\text{sn}(u,k) = \frac{1}{k}\text{sn}(ku,\frac{1}{k})$, $\text{cn}(u,k) = \text{dn}(ku,\frac{1}{k})$, $\text{dn}(u,k) = \text{cn}(ku,\frac{1}{k})$ \cite{ByrdFriedman54}.
The motion of the center of mass can be recovered straightforwardly using the angle $\theta$ to bring the body-frame velocity $\bm{U}$ into a nonrotating frame. 

In writing the forms \eqref{thetapendulum} and \eqref{solb}, we arranged for a positive restoring term on the pendulum, and we pulled out the larger of two mass coefficients into the ratio $\gamma$ to create two nondimensional positive quantities.  This is the mass coefficient corresponding to the flutter axis, that is, the second label in the flutter subscript. 
Because of these choices, the solutions for the other axes are not simply cyclic permutations. 
For rotation around the $c$ axis, one may write 
$P_a = P\text{sn}(\eta t, k)$ and $P_b= P\text{cn}(\eta t, k)$ and perform the substitutions $M_c \rightarrow M_b$ and $I_b \rightarrow I_c$.
For rotation around the $a$ axis, one may write 
$P_b = P\text{sn}(\eta t, k)$ and $P_c= P\text{cn}(\eta t, k)$ and perform the substitutions $M_a \rightarrow M_b$ and $I_b \rightarrow I_a$.

\subsection{Linear stability of planar modes}\label{planarmodestability}

Again let the components consist of a base state and a perturbed field, denoted $P_a = P_{a0} + p_a$, and so on, with the base state now a time-dependent planar solution from Section \ref{planarmodes} above. 
Linearizing \eqref{kirchcomp} about such a solution results in two decoupled three-dimensional systems describing the ``in plane'' and ``out of plane'' parts of the perturbed dynamics. The former type of perturbation simply shifts the system to another orbit on the integrable submanifold, while the latter type tells us about stability of the full six-dimensional system describing motion of a body in three spatial dimensions. 
For example, for $b$ axis tumbling or fluttering modes, the nonvanishing terms in the base state are time-dependent solutions $P_{a0}$, $P_{c0}$, $L_{b0}$ taken from \eqref{solb}, the uninteresting ``in plane'' system is
\begin{equation}
\begin{bmatrix}
    \dot p_a \\
    \dot p_c \\ 
    \dot l_b   
\end{bmatrix}= 
\begin{bmatrix}
   0 & -\frac{1}{I_b} L_{b0}  & -\frac{1}{I_b} P_{c0} \\ 
   \frac{1}{I_b} L_{b0}  & 0 & \frac{1}{I_b} P_{a0}  \\
   (\frac{1}{M_c}-\frac{1}{M_b}) P_{c0}  & (\frac{1}{M_c}-\frac{1}{M_b}) P_{a0}   & 0
\end{bmatrix}
\begin{bmatrix}
    p_a  \\ 
    p_c  \\
    l_b  
\end{bmatrix} \, , \label{eq:inplane}
\end{equation}
and the important ``out of plane'' system is
\begin{equation}
\begin{bmatrix}
    \dot p_b \\
    \dot l_a \\ 
    \dot l_c   
\end{bmatrix}=
\begin{bmatrix}
   0& \frac{1}{I_a} P_{c0}  & -\frac{1}{I_c} P_{a0}  \\ 
   (\frac{1}{M_b}-\frac{1}{M_a}) P_{c0}  & 0  &  (\frac{1}{I_b}-\frac{1}{I_a}) L_{b0}  \\ 
   (\frac{1}{M_a}-\frac{1}{M_c}) P_{a0}  & (\frac{1}{I_a}-\frac{1}{I_c}) L_{b0}  & 0
\end{bmatrix}
\begin{bmatrix}
    p_b  \\ 
    l_a \\ 
    l_c   
\end{bmatrix} \, . \label{eq:outofplane}
\end{equation}
Systems of equations corresponding to the other axes are obtained by cyclic permutation. 

To ascertain stability from the system \eqref{eq:outofplane} with its time-periodic matrix, we use the ``practical numerical procedure'' for applying Floquet theory outlined in \cite[section 3.1.2]{kovacic2018mathieu}, which involves numerical integration for one period. The periodicity depends on $\eta$ and 
 $k$ for the base state, and for each such state we integrate for a time $\tau$ expressible in terms of the complete elliptic integral of the first kind, 
\begin{equation}
    \eta \tau 
    =4\int^{\frac{\pi}{2}}_{0} \frac{dv}{\sqrt{1-k^2\sin^2v}} \, . 
\end{equation}
We construct a constant matrix $\text{C}$ whose columns are the states at time $\tau$ resulting from the evolution of unit perturbations of each of the three quantities. 
The determinant of this matrix will be unity, because the system \eqref{eq:outofplane} is traceless. 
Instability corresponds to the presence of any eigenvalues $\ell$ of this matrix with norm greater than unity.  These are obtained by solving the  characteristic equation in the form
\begin{equation}
	\ell^3-\text{Tr}(\text{C})\ell^2 + \tfrac{1}{2}\left[(\text{Tr}(\text{C}))^2-\text{Tr}(\text{C}^2)\right]\ell - 1=0 \, . \label{cubic}
\end{equation}
One eigenvalue will be unity, and the other two will either be real or a complex conjugate pair \cite{guckenheimer2013nonlinear}, depending on the sign of the discriminant of the cubic \eqref{cubic}. 
Because the product of eigenvalues, the determinant, is unity, the complex conjugate pair have unit norm, and the system is marginally stable.  And if the eigenvalues are real, one is larger than unity and the system is unstable \cite{kovacic2018mathieu}. 
We express our results in terms of the closely related matrix that can be obtained by taking a matrix logarithm \cite{Chicone}.  Its eigenvalues $\lambda$ have the relation 
\begin{equation}
	\tau\lambda = \ln \ell \, , 
\end{equation} 
and thus instability corresponds to any $\lambda$ having positive real part.  For the present system, either $\text{Re}(\lambda)=0$ or $\text{Re}(\lambda)=\lambda$. 

\begin{figure}[h]
    \centering
    \includegraphics[width=6in]{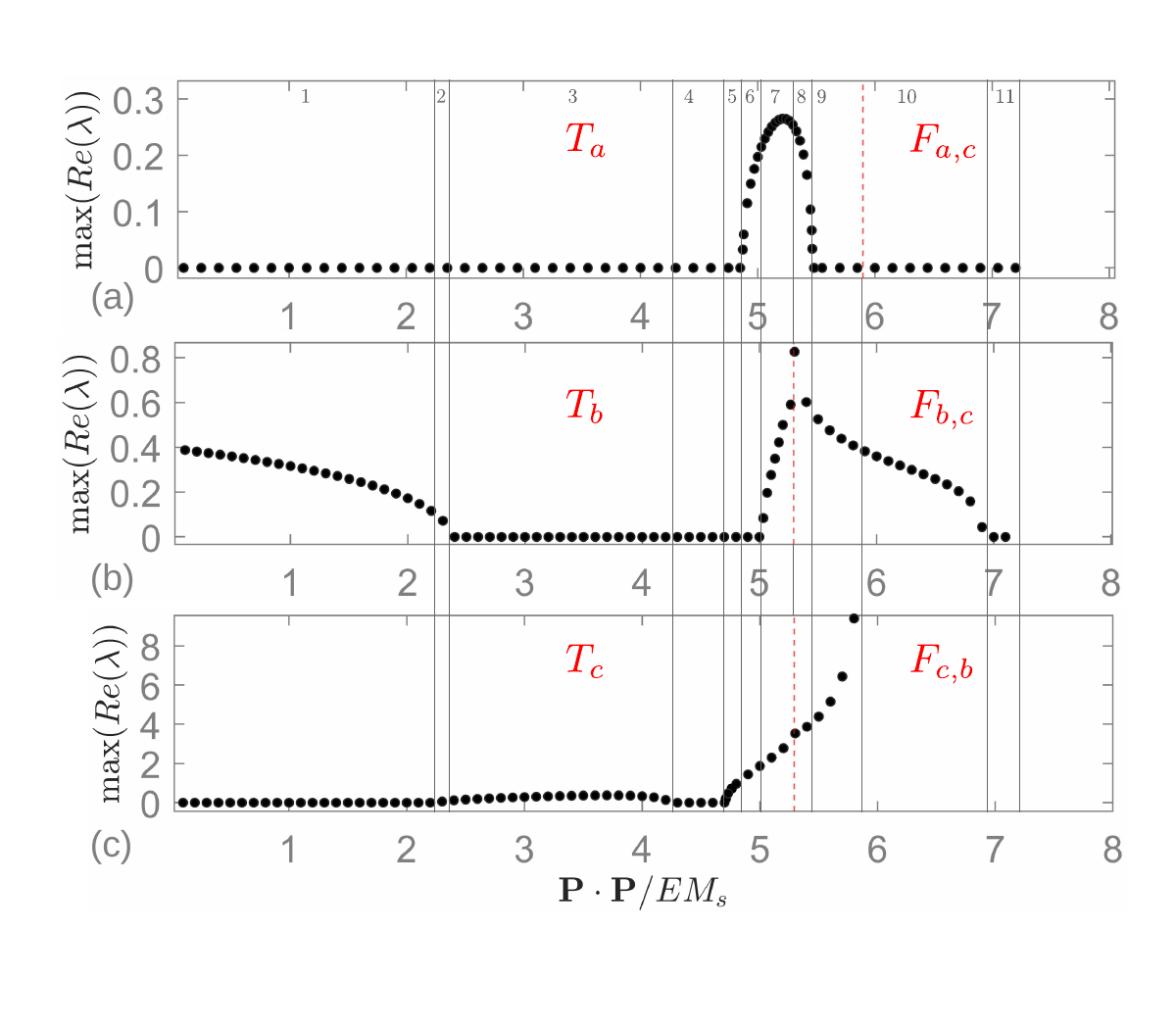}
    \vspace{-1.5cm}
    \caption{The largest real part of any eigenvalue for the perturbed dynamics around planar tumbling T and fluttering F around the $a$, $b$, $c$ axes (labeled accordingly), in terms of the linear momentum to energy ratio $\frac{\bm{P} \cdot \bm{P}}{EM_s}$ normalized using the total ellipsoid mass $M_s$.  Nonzero values indicate instability.  Eleven zones are indicated and separated by vertical lines, according to the existence and stability of the six possible modes.  Red dashed lines indicate transitions between tumble and flutter.  The modes $\text{F}_{b,c}$ and $\text{F}_{c,b}$ appear at the same ratio.  The appearance of mode $\text{F}_{a,c}$ coincides with the  disappearance of $\text{F}_{c,b}$, at the upper limit of the ratio for $c$ axis modes.  The upper limit for $a$ and $b$ axis modes coincides at the end of zone 11.}
    \label{bifurcation}
\end{figure}

The stability of tumbling and fluttering around the three axes is shown in Figure \ref{bifurcation}, using an isotropic nondimensional ratio of linear momentum and energy $\frac{\bm{P} \cdot \bm{P}}{EM_s}$ with the total ellipsoid mass $M_s$ in the denominator.  This diagram contains eleven zones, according to the existence and stability of the six possible planar modes.  The linear stability determined by Floquet analysis of the three-dimensional ``out of plane'' perturbations was confirmed by numerical integration of the full six-dimensional system. 

\subsection{Map of landmark solutions}\label{map}

Pure and planar modes are shown schematically in the six-dimensional $P$-$L$ space in Figure \ref{map}.   Families of such solutions concentrically fill the space, but only one example of each kind is shown.  Solutions consist of a single connected piece in each space, but come in pairs.  Pure modes are fixed points; each $P$ point can be associated with one of a pair of $L$ points and \emph{vice versa}.  A tumbling solution traces out a circle in $P$ space and one of a pair of corresponding line segments in $L$ space that do not pass through the origin.  A fluttering solution traces out one of a pair of circular arcs in $P$ space corresponding to a line segment in $L$ space that passes through the origin. 
\begin{figure}[!h]
    \centering
    \includegraphics[width=5in]{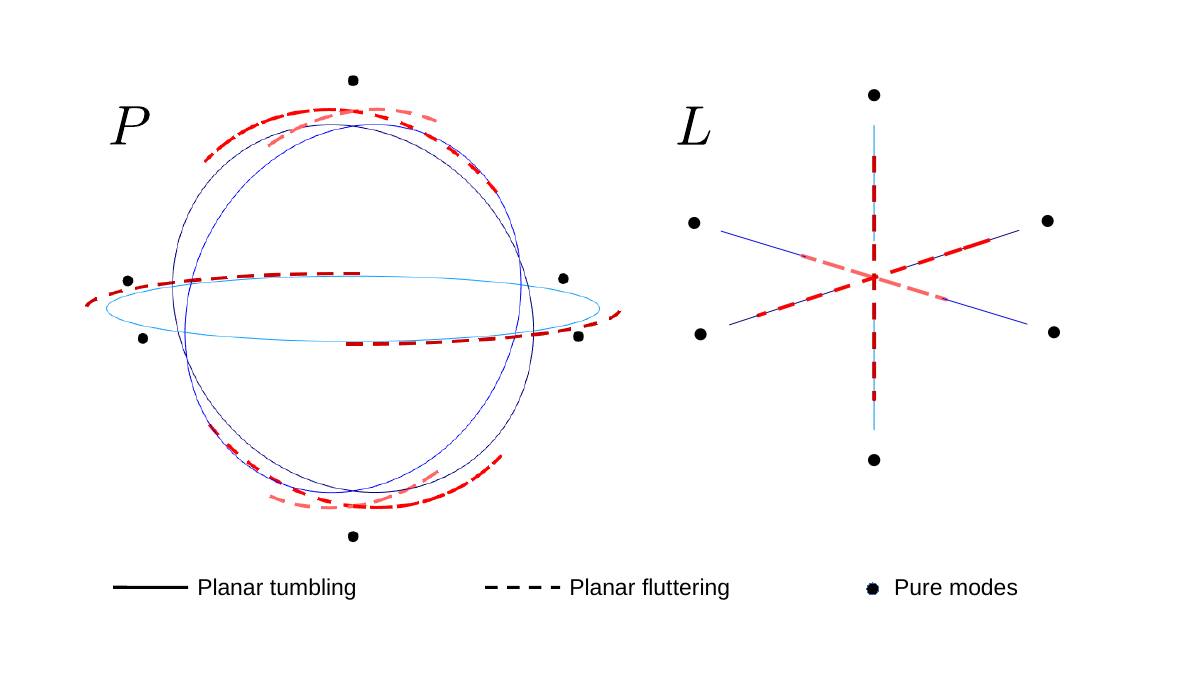}
    \vspace{-0.5cm}
    \caption{Map of integrable landmark solutions in $P$-$L$ space, showing one example of each kind.  Pure modes (black dots), tumbling (solid lines in dark, medium, or light blue), and fluttering (dashed lines in dark, medium, or light red). Solutions consist of a single connected piece in each space, but come in pairs.}
    \label{map}
\end{figure}

\clearpage

\section{Perturbations of unstable landmark solutions}\label{examplesolutions}

In this section, we present a selection of behaviors that approach close to one or more unstable landmark solutions whose form in momentum space is indicated in Figure \ref{map}.  

\subsection{Unstable pure modes}\label{perturbpure}

In this subsection, we refer to unstable regions in Figure \ref{PmodeB} in Section \ref{puremodestability}. Near pure modes, $\bm{L} \cdot \bm{P}$ is close to maximal.  

\begin{figure}[h]
    \centering
    \includegraphics[width=6in]{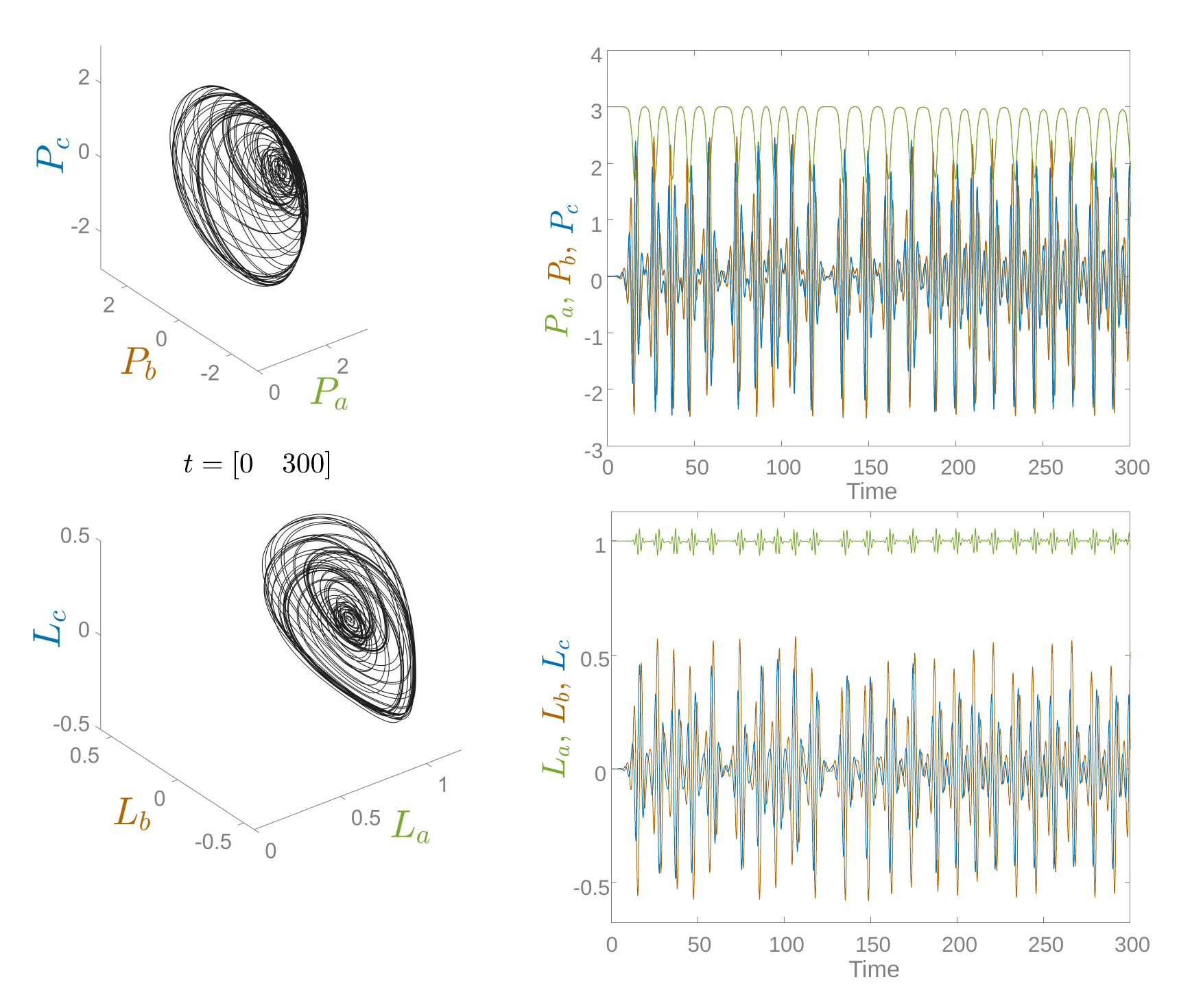}
    \vspace{-0.5cm}
    \caption{Twirling around an unstable $a$-axis pure mode. Trajectory in momentum space, and time series of the momentum components. The base state has $\frac{L_a}{P_a}=\frac{1}{3}$.   Initial conditions $\bm{P}=[3.0001,0.0001,0.0001]$, $\bm{L}=[1.0001,0.0001,0.0001]$, with $\bm{P} \cdot \bm{ P}=9.0006$, $\bm{L} \cdot \bm{ P}=3.0004$, $E=2.8710$. }
    \label{Puremode1}
\end{figure}

\begin{figure}[h]
    \centering
    \includegraphics[width=6in]{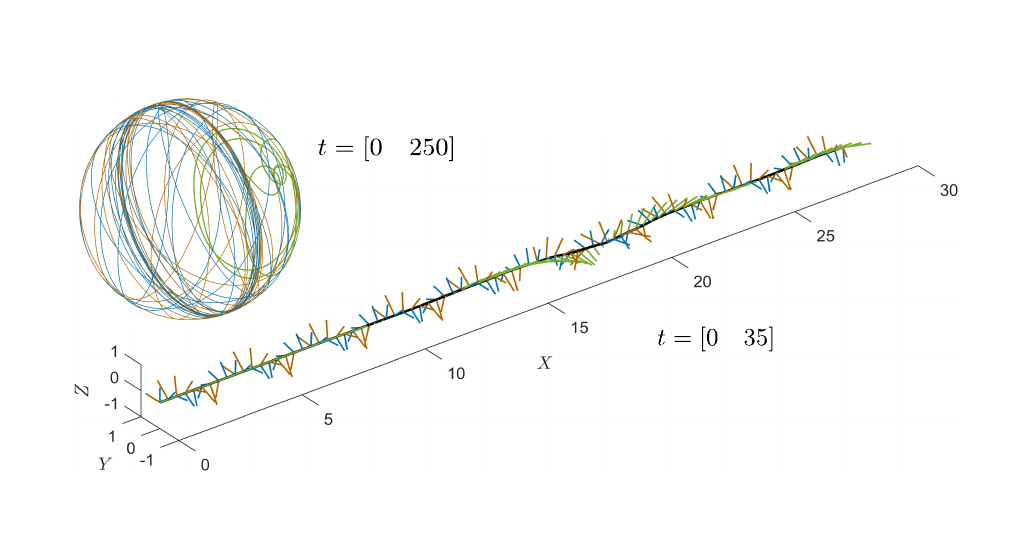}
    \vspace{-1cm}
    \caption{Twirling around an unstable $a$-axis pure mode, with conditions as in Figure \ref{Puremode1}. 
    Evolution of the body axes, scaled according to body dimensions, along a center of mass trajectory in real space, and of the body frame on a unit sphere. Axes $a, b, c$ are in green, brown, blue. Note different time intervals.}
    \label{COMpuremode1}
\end{figure}

The $a$-axis pure mode is observed to be unstable to a twirling motion, associated with a homoclinic connection previously documented by Holmes and co-workers \cite{Holmes98}, who called it ``wobbling''.  Excursions away from the unstable state repeat in an irregular manner.  Figure \ref{Puremode1} shows a trajectory in momentum space, and time series of the momentum components.  Figure \ref{COMpuremode1} shows the evolution of the body axes on a center of mass trajectory in real space, and of the body frame on a unit sphere. The center of mass trajectory is a complex three-dimensional motion, not easily seen when plotting with aspect ratio unity as here. 

\begin{figure}[h]
    \centering
    \includegraphics[width=6in]{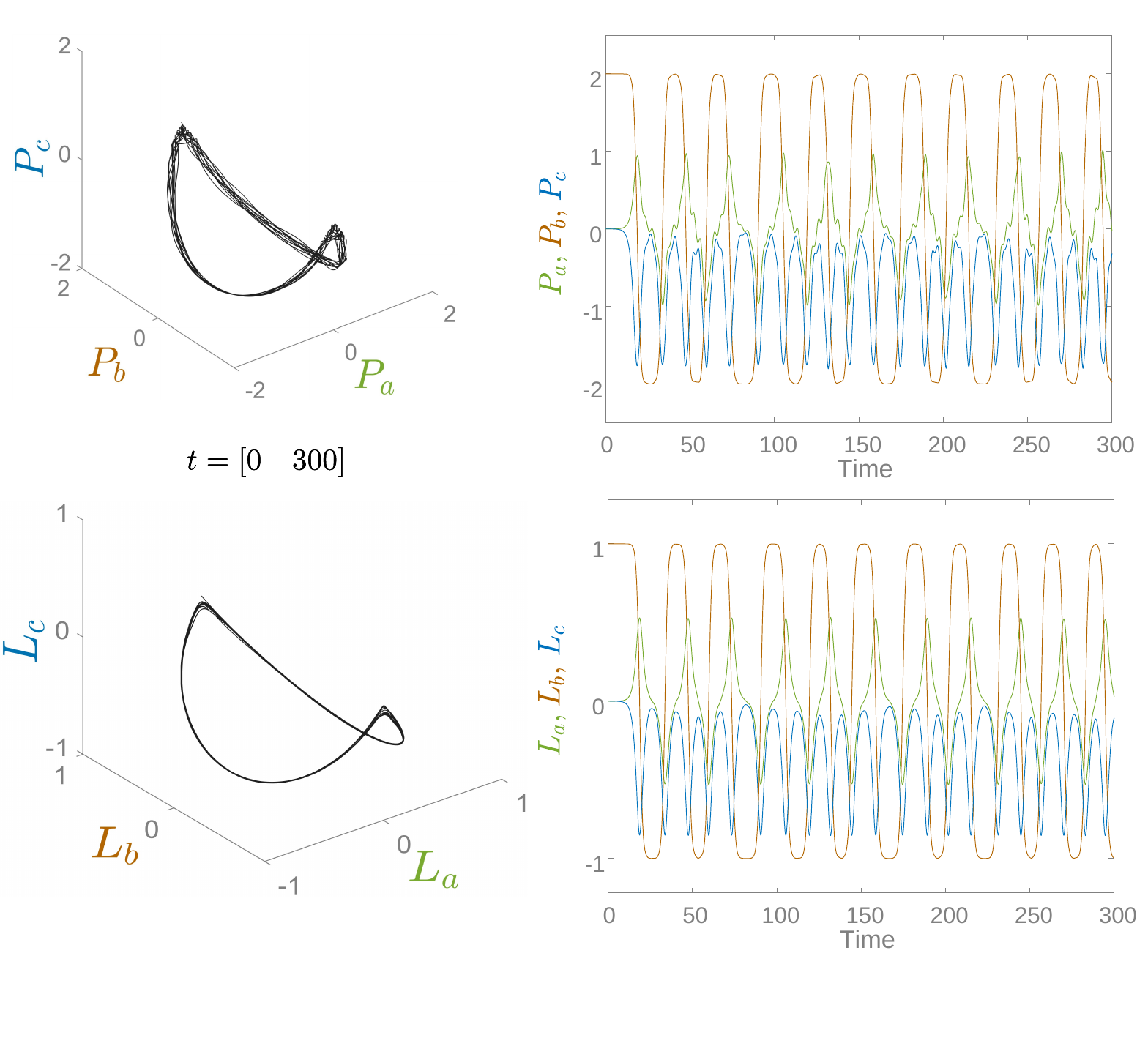}
    \vspace{-1cm}
     \caption{Flipping between positive and negative orientation unstable $b$ axis pure modes.  Trajectory in momentum space, and time series of the momentum components.  The base state has $\frac{L_b}{P_b}= \frac{1}{2}$.  Initial conditions $\bm{P}=[0.0001,2.0001,0.0001]$, $\bm{L}=[0.0001,1.0001,0.0001]$, with $\bm{P} \cdot \bm{ P}=4.0004$, $\bm{L} \cdot \bm{ P}=2.0003$, $E=1.4623$. }
    \label{mode2B}
\end{figure}

The $b$-axis pure mode is observed to be unstable to flipping between positive and negative orientation, associated with heteroclinic connections. An example of motion of this type is shown in Figure \ref{mode2B}.  The body flips through regular alternation between two paths, for which the $c$ components have the same sign and the $a$ components have opposite signs. 

\clearpage

\subsection{Unstable planar modes}\label{perturbplanar}

In this subsection, we refer to unstable regions corresponding to numbered zones in Figure \ref{bifurcation} in Section \ref{planarmodestability}. 
Near planar modes, $\bm{L} \cdot \bm{P}$ is close to zero.  Recall that this does not imply planarity of motions. There is a rich chaotic sea of solutions with $\bm{L} \cdot \bm{P}=0$, and perturbations of unstable planar solutions will travel very far from planarity. 
Most of the perturbations shown here are generic, but two of them maintain $\bm{L} \cdot \bm{P}$ strictly zero. 
We probe unstable behavior both relatively near and far away from the bifurcations separating the zones.  As one might expect, we tend to observe greater propensity for chaos farther from these bifurcations. 

\begin{figure}[h]
    \centering
    \includegraphics[width=6in]{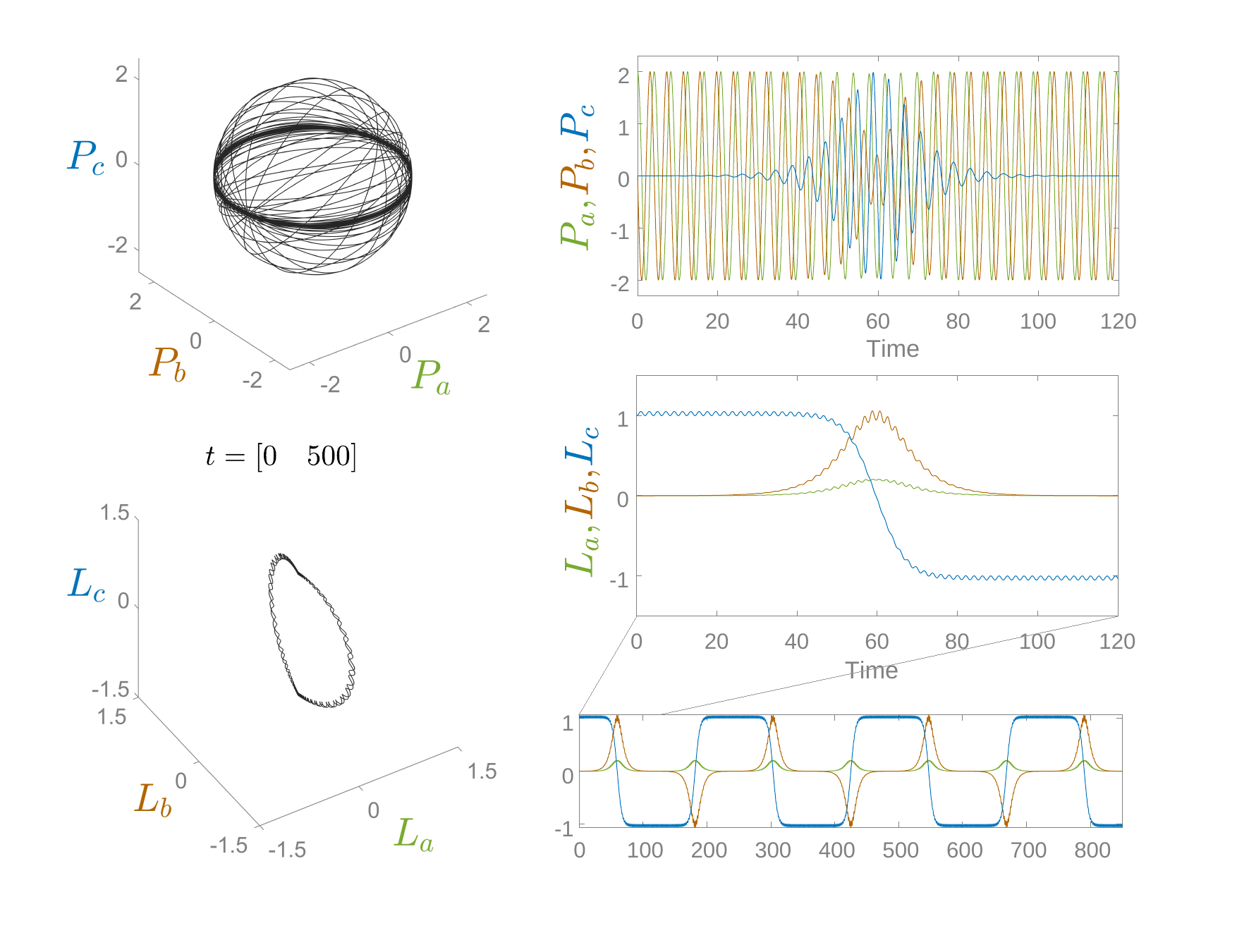}
    \vspace{-1cm}
    \caption{Regular flipping between positive and negative orientation unstable $c$ axis tumbling modes $\mathbf{T_c}$.   Trajectory in momentum space, and time series of the momentum components, including a longer time sequence for angular momentum.  Note different time intervals. Initial conditions $\bm{P}=[2,0,0.0001]$, $\bm{L}=[0.0001,0.0001,1]$, with $\bm{P} \cdot \bm{P}=4$, $\bm{L} \cdot \bm{P}=0.0003$, $E=1.5$, $\frac{\bm{P} \cdot \bm{P}}{EM_s}=2.6667$ (Zone 3).}
    \label{tcflipreg}
\end{figure}

\begin{figure}[h]
    \centering
    \includegraphics[width=6in]{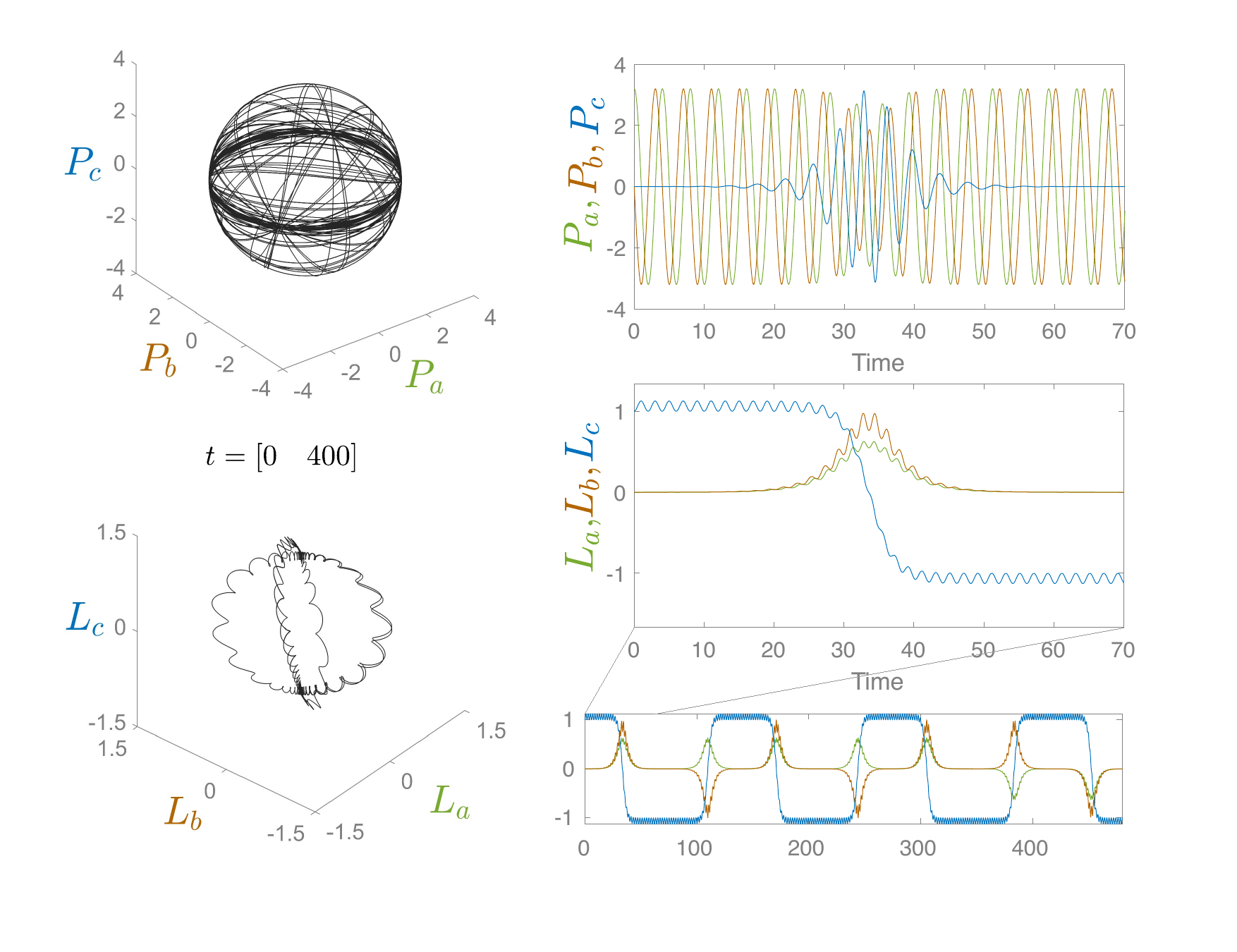}
    \vspace{-1cm}
    \caption{Chaotic flipping between positive and negative orientation unstable $c$ axis tumbling modes $\mathbf{T_c}$.   Trajectory in momentum space, and time series of the momentum components. Note different time intervals. Initial conditions $\bm{P}=[3.2,0,0.0001]$, $\bm{L}=[0.0001,0.0001,1]$, with $\bm{P} \cdot \bm{P}=10.2$, $\bm{L} \cdot \bm{P}=0.000042$, $E=2.6$, $\frac{\bm{P} \cdot \bm{P}}{EM_s}=3.9231$ (Zone 3).}
    \label{tcflipchao}
\end{figure}

\begin{figure}[h]
    \centering
    \includegraphics[width=6in]{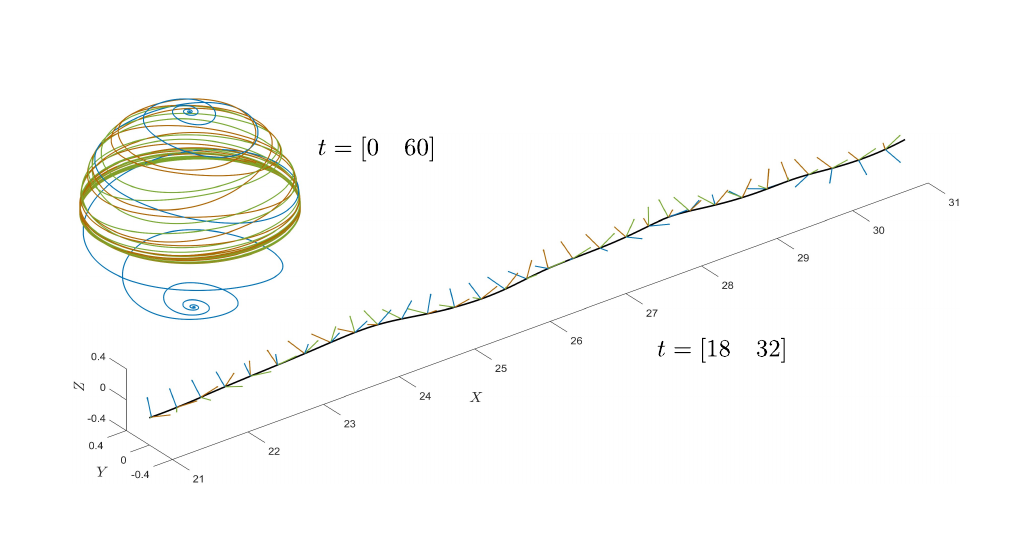}
    \vspace{-1cm}
    \caption{
    A single flip between positive and negative orientation unstable $c$ axis tumbling modes $\mathbf{T_c}$, with conditions as in Figure \ref{tcflipreg}. 
    Evolution of the body axes, scaled according to body dimensions,  along a center of mass trajectory in real space, and of the body frame on a unit sphere. 
    Axes $a, b, c$ are in green, brown, blue. Note different time intervals.}
    \label{tcflipregspace}
\end{figure}

In Zone 3, only the $\text{T}_c$ mode is unstable, to flipping between positive and negative orientation.  
Figure \ref{tcflipreg} is an example of regular flipping for a body near the bifurcation between Zones 2 and 3.  The two $\text{T}_c$ modes correspond to very short line segments in $L$ space away from the origin, and a circle in $P$ space. The body travels between these by regularly alternating between two branches whose $L_a$ components have the same sign and $L_b$ components have opposite signs, oscillating within a heart-shaped region. 
Figure \ref{tcflipchao} is deeper within Zone 3, with a more unstable eigenvalue. The body travels between the two $\text{T}_c$ modes by irregular switching between four branches corresponding to two possible signs each of $L_a$ and $L_b$ components.  Initial conditions differing at the fourth decimal place are seen to follow different flipping sequences. 
The orientation of the body during a flip is shown in Figure \ref{tcflipregspace}. 
The $c$ axis takes a spiral path from one pole to the other, while the other axes return to the same circle. Again, the center of mass trajectory is three-dimensional. 
Similarly, in Zone 1, only the $\text{T}_b$ mode is unstable, to flipping between positive and negative orientation. We note that within Zones 1 and 3, when perturbations were chosen such that $\bm{L} \cdot \bm{P}$ remained strictly zero, only regular two-branch flipping was observed.  

\clearpage

\begin{figure}[h]
    \centering
    \includegraphics[width=6in]{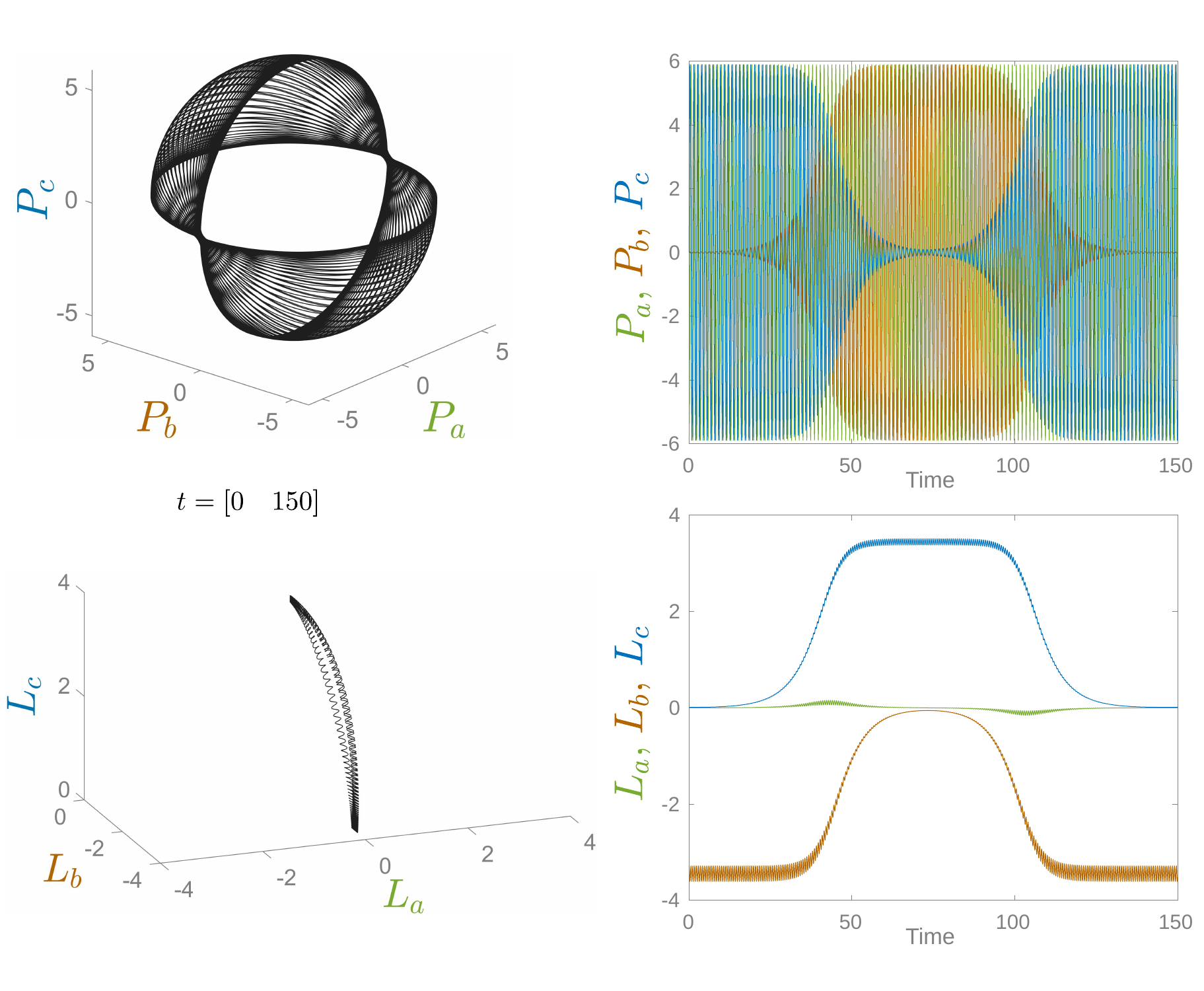}
    \vspace{-0.5cm}
    \caption{Motion associated with a homoclinic connection of an unstable $b$ axis tumbling mode $\text{T}_b$ that approaches close to an unstable $c$ axis tumbling mode $\text{T}_c$. 
    Trajectory in momentum space, and time series of the momentum components. 
     Initial conditions $\bm{P}=[-5.8916,0,0]$, $\bm{L}=[0,-3.2836,0.01]$, with $\bm{P} \cdot \bm{P}=34.7117$, $\bm{L} \cdot \bm{P}=0$, $E=15$, $\frac{\bm{P} \cdot \bm{P}}{EM_s}=2.3141$ (Zone 2).}
    \label{tbhomo}
\end{figure}

In Zone 2, both the $\text{T}_b$ and $\text{T}_c$ modes are unstable.  Several types of motion are observed, to be detailed in the following Section \ref{fillspace}. Figure \ref{tbhomo} is an example of an orbit associated with a homoclinic connection of a $\text{T}_b$ mode, but one that approaches close to a $\text{T}_c$ mode.  These modes correspond to very short line segments in $L$ space away from the origin, and circles in $P$ space.  For this example, the perturbation was chosen to retain the condition of vanishing $\bm{L} \cdot \bm{P}$.  We will see in the following Section \ref{fillspace} that another bifurcation affecting connectivity, rather than stability, occurs in this zone, at least for vanishing $\bm{L} \cdot \bm{P}$.  
 
\begin{figure}[h]
    \centering
    \includegraphics[width=6in]{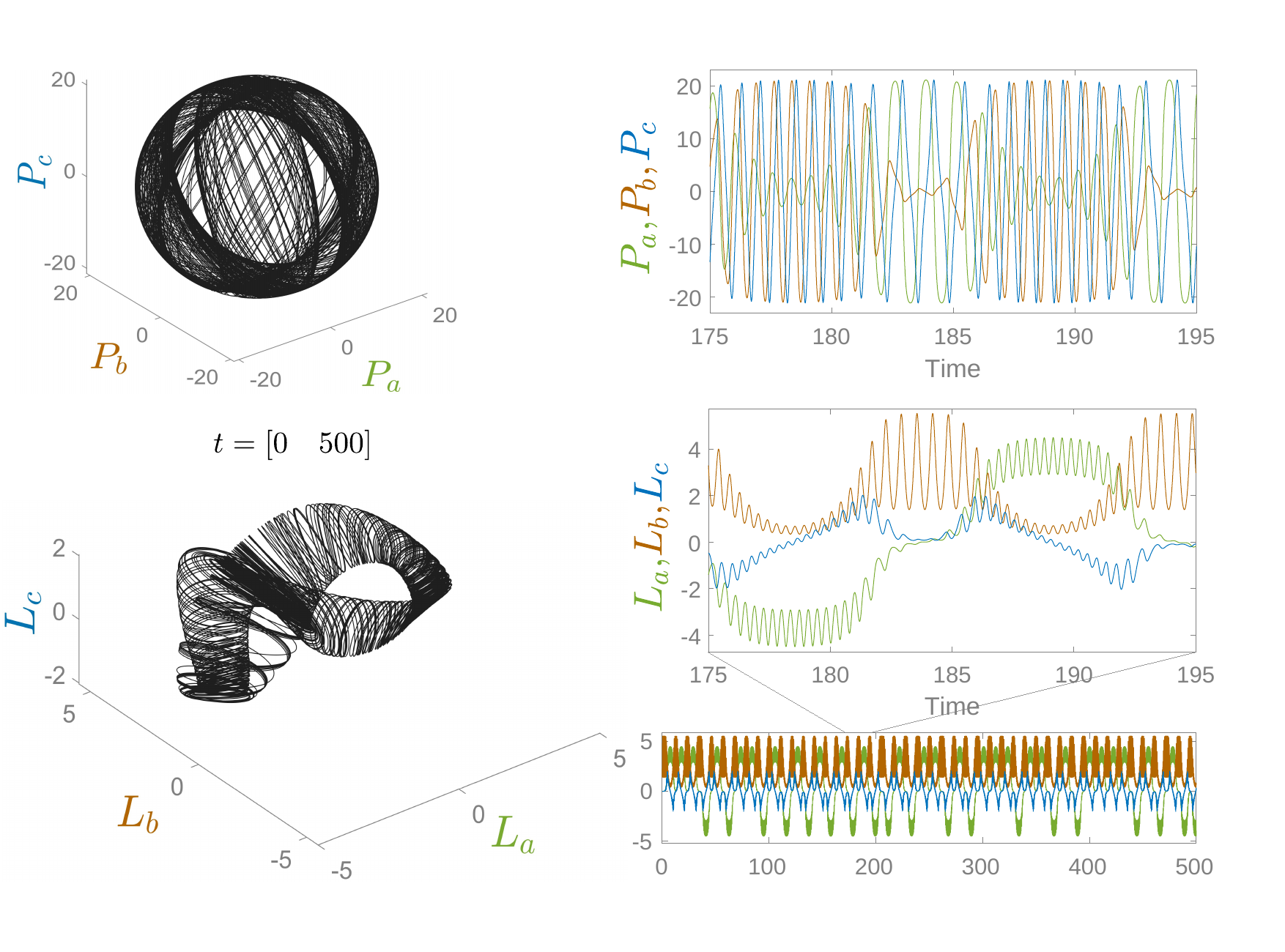}
    \vspace{-0.5cm}
    \caption{Chaotic switching in a figure-eight around an unstable $b$ axis tumbling mode $\text{T}_b$ that approaches close to positive and negative orientation unstable $a$ axis tumbling modes $\text{T}_a$. Trajectory in momentum space, and time series of the momentum components, including a longer time sequence for angular momentum.    Initial conditions $\bm{P}=[21.1,0.001,0.001]$, $\bm{L}=[0.001,1.4,0.001]$, with $\bm{P} \cdot \bm{P}=445.21$, $\bm{L} \cdot \bm{P}=0.0225$, $E=85.6795$, $\frac{\bm{P} \cdot \bm{P}}{EM_s}=5.1962$ (Zone 7).}
    \label{tbpmta}
\end{figure}

\begin{figure}[h]
    \centering
    \includegraphics[width=6in]{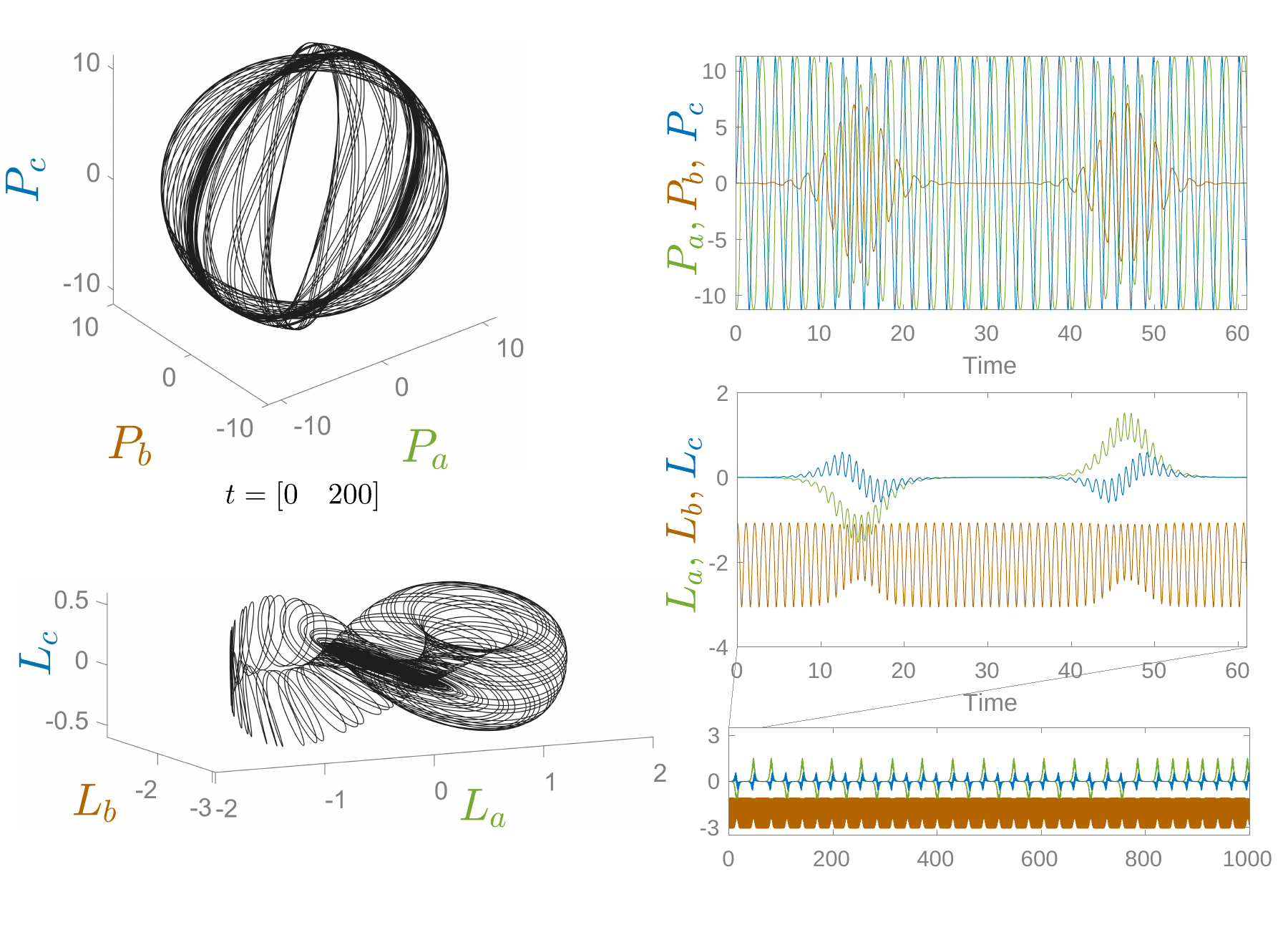}
    \vspace{-0.5cm}
    \caption{Chaotic switching in a figure-eight around an unstable $b$ axis tumbling mode $\text{T}_b$. Trajectory in momentum space, and time series of the momentum components, including a longer time sequence for angular momentum. Note different time intervals. 
    Initial conditions $\bm{P}=[-11.2916,0,0]$, $\bm{L}=[0,1.0734,0.01]$, with $\bm{P} \cdot \bm{P}=127.5$, $\bm{L} \cdot \bm{P}=0$, $E=25$, $\frac{\bm{P} \cdot \bm{P}}{EM_s}=5.1$ (Zone 7).}
    \label{doublebhomo}
\end{figure}

\begin{figure}[h]
    \centering
    \includegraphics[width=6in]{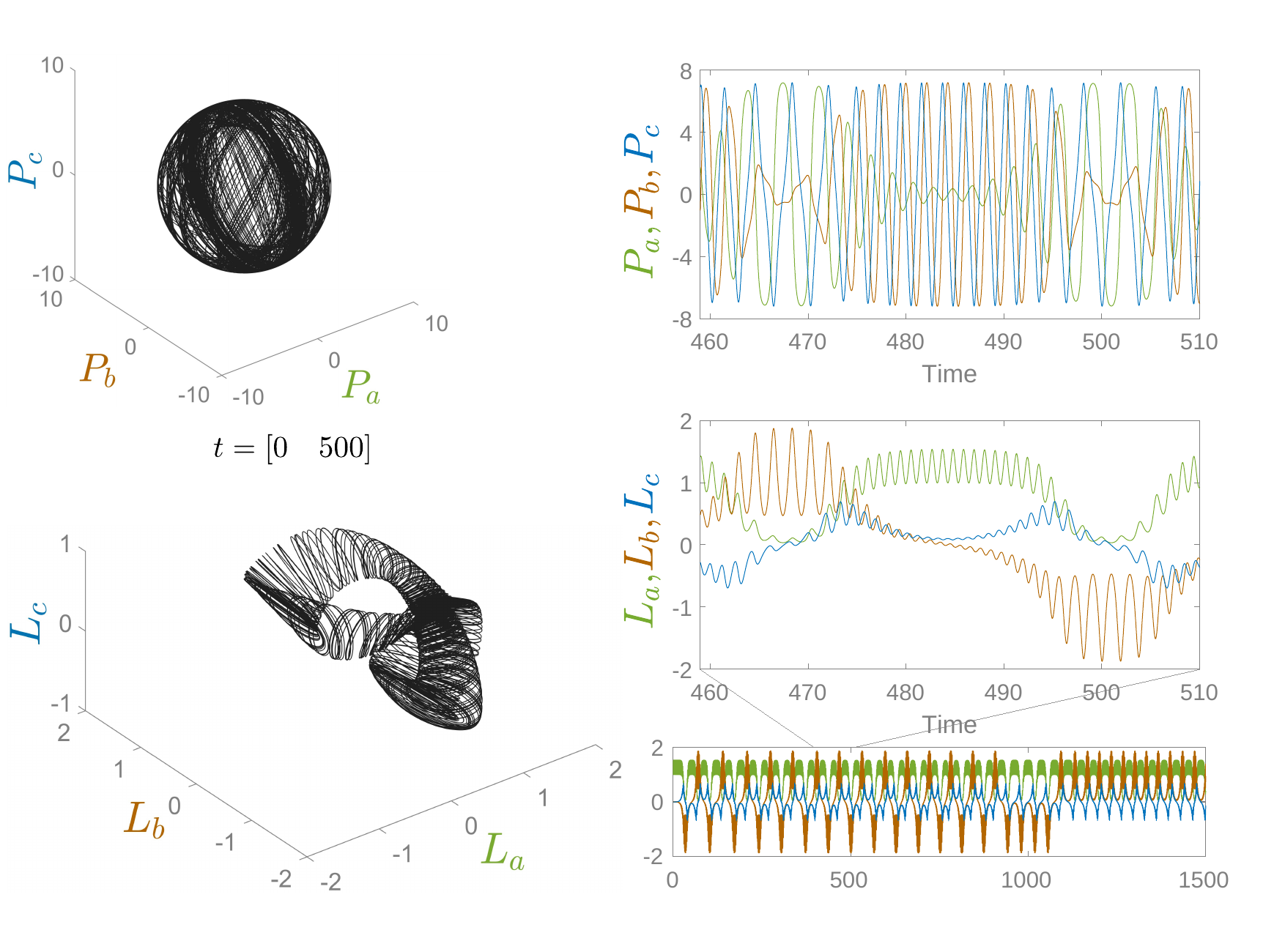}
    \vspace{-0.5cm}
    \caption{Chaotic switching in a figure-eight around an unstable $a$ axis tumbling mode $\text{T}_a$ that approaches close to positive and negative orientation unstable $b$ axis tumbling modes $\text{T}_b$. Trajectory in momentum space, and time series of the momentum components, including a longer time sequence for angular momentum. Note different time intervals. 
    Initial conditions $\bm{P}=[0.0011,7.2,0.001]$, $\mathbf{L}=[1,0.001,0.0001]$, with $\bm{P} \cdot \bm{ P}=51.84$, $\bm{L} \cdot \bm{P}=0.0083$, $E=9.9$, $\frac{\bm{P} \cdot \bm{P}}{EM_s}=5.2364$ (Zone 7).}
    \label{tapmtb}
\end{figure}

\begin{figure}[h]
    \centering
    \includegraphics[width=6in]{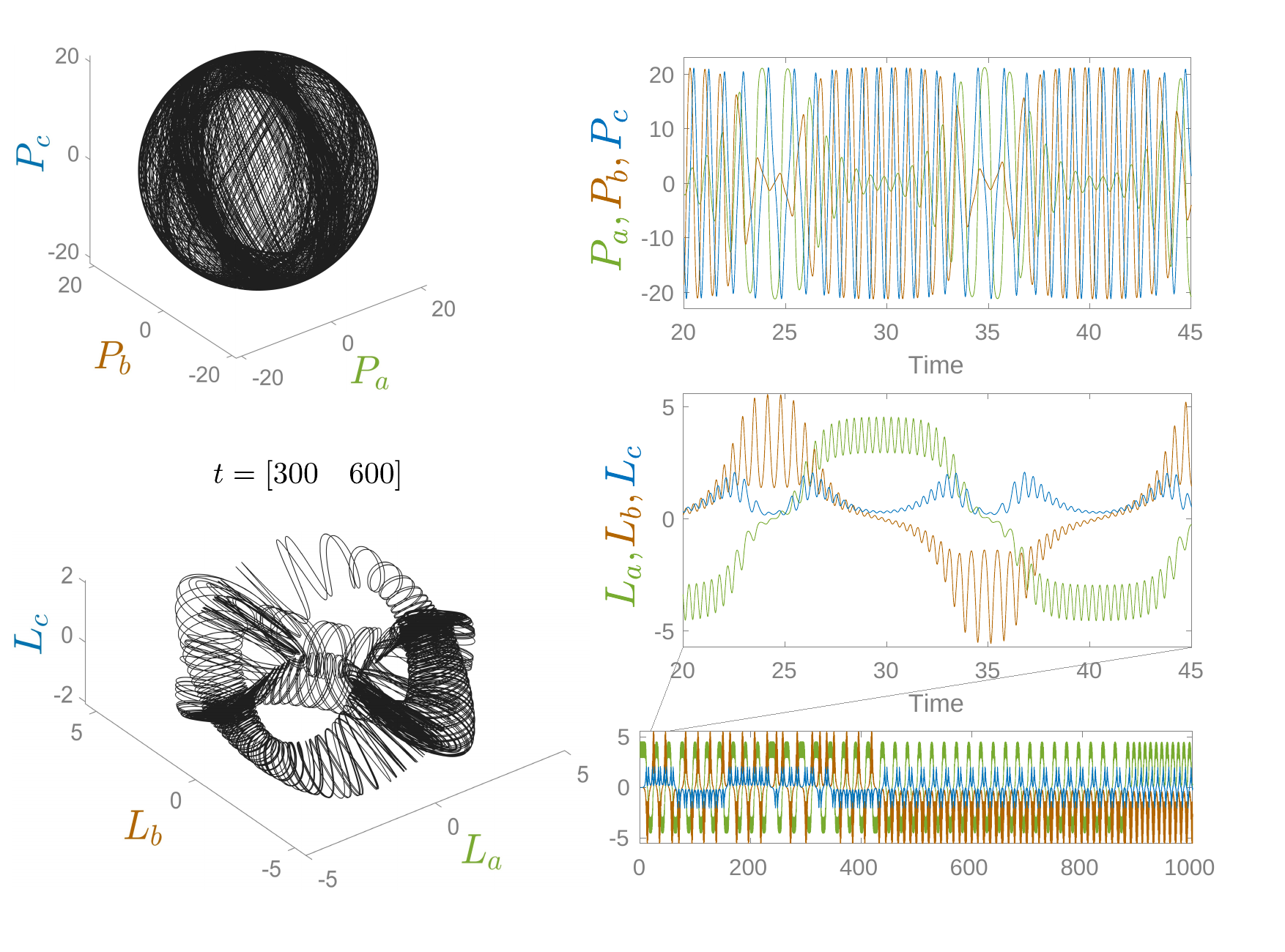}
    \vspace{-0.5cm}
    \caption{Chaotic switching between positive and negative orientation unstable $a$ and $b$ axis tumbling modes $\text{T}_a$ and $\text{T}_b$. Trajectory in momentum space, and time series of the momentum components, including a longer time sequence for angular momentum. Note different time intervals.     Initial conditions $\bm{P}=[0.001,21.2,0]$, $\bm{L}=[2.9545,0.001,0.001]$, with $\bm{P} \cdot \bm{P}=449.44$, $\bm{L} \cdot \bm{P}=0.0242$, $E=86.4790$, $\frac{\bm{P} \cdot \bm{P}}{EM_s}=5.1971$ (Zone 7).}
    \label{pmtapmtb2}
\end{figure}

In Zone 7, all three tumbling modes are unstable, and a variety of motions are observed, to be detailed in the following Section \ref{fillspace}. Figures \ref{tbpmta} and \ref{doublebhomo} both show qualitatively similar chaotic orbits associated with a figure-eight trajectory starting near a $\text{T}_b$ mode and approaching positive and negative orientation $\text{T}_a$ modes through irregular switching, with two branches for each orientation. 
The modes correspond to line segments in $L$ space away from the origin, and circles in $P$ space.  In the example in Figure \ref{tbpmta}, the system lingers longer near the positive and negative orientation $\text{T}_a$ modes.  Similar trajectories are observed that are associated with flipping between these modes such that there is close approach to a $\text{T}_b$ mode. 
The perturbation is generic in Figure \ref{tbpmta}, but is chosen to retain the condition of vanishing $\bm{L} \cdot \bm{P}$ in Figure \ref{doublebhomo}. 
Figure \ref{tapmtb} shows a trajectory analogous to that in Figure \ref{tbpmta} with the axes switched.   Here the body begins near a $\text{T}_a$ mode and approaches positive and negative orientation $\text{T}_b$ modes.  Flipping between positive and negative orientation $\text{T}_b$ modes are also observed.  
We will see in the following Section \ref{fillspace} that a bifurcation in connectivity occurs in this zone as well, at least for vanishing $\bm{L} \cdot \bm{P}$, which is related to the different behaviors in these two figures. 
Figure \ref{pmtapmtb2} shows a more complex motion in which the system visits, through chaotic switching, four saddles corresponding to the two possible orientations of $\text{T}_a$ and $\text{T}_b$, with each different pair connected by a pair of branches. 

\clearpage

\begin{figure}[h]
    \centering
    \includegraphics[width=6in]{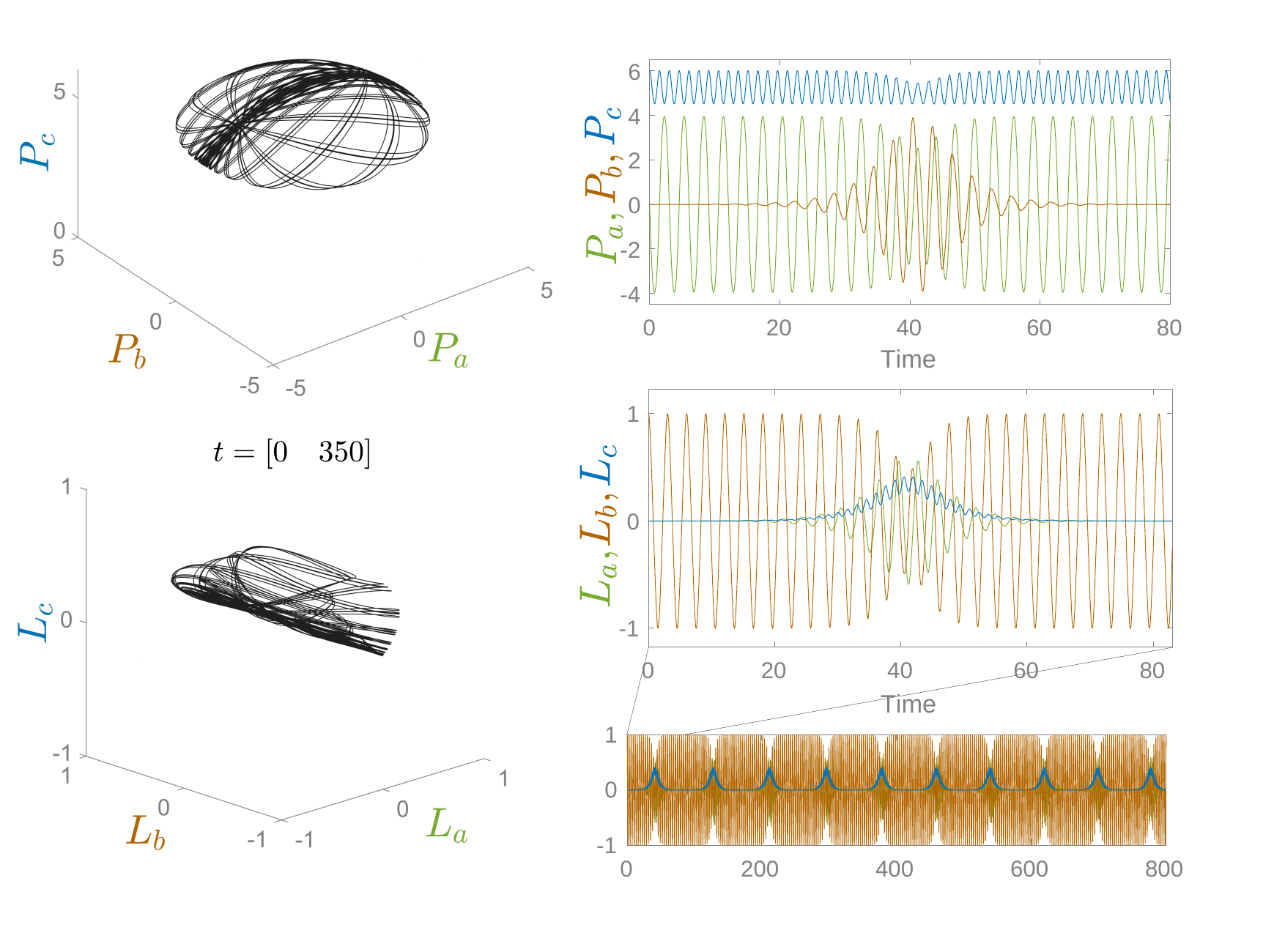}
    \vspace{-0.5cm}
    \caption{Regular spinning around an unstable $b$ axis fluttering mode $\text{F}_{b,c}$. 
    Trajectory in momentum space, and time series of the momentum components, including a longer time sequence for angular momentum.  Note different time intervals. Initial conditions $\bm{P}=[0,0.0001,6]$, $\bm{L}=[0.0001,1,0.0001]$, with $\bm{P} \cdot \bm{P}=36$, $ \bm{L} \cdot \bm{P}=0.0007$, $E=5.7752$, $\frac{\bm{P} \cdot \bm{P}}{EM_s}=6.2336$ (Zone 10). }
    \label{flutterregular}
\end{figure}

\begin{figure}[h]
    \centering
    \includegraphics[width=6in]{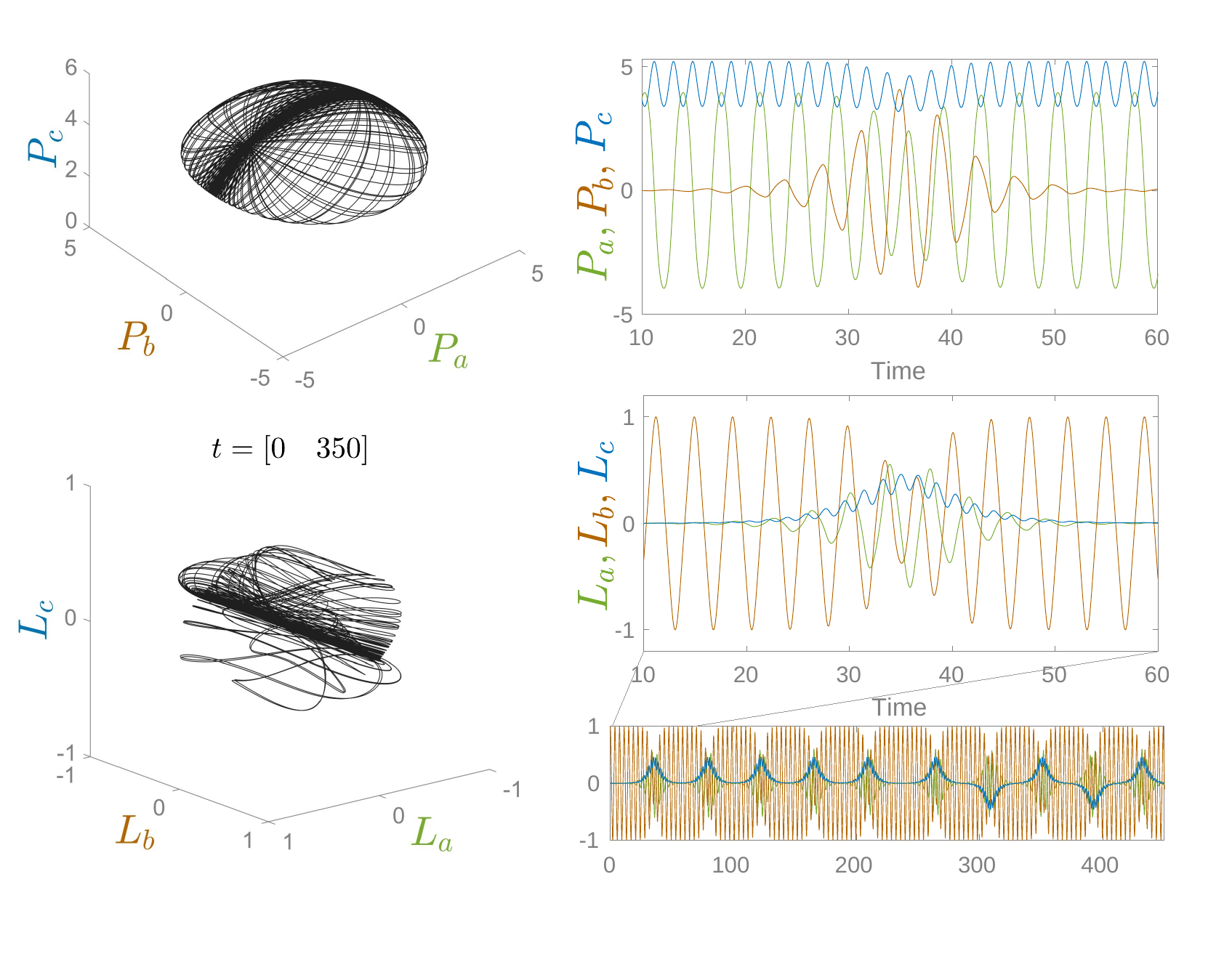}
    \vspace{-1cm}
    \caption{Chaotic spinning around an unstable $b$ axis fluttering mode $\text{F}_{b,c}$. 
    Trajectory in momentum space, and time series of the momentum components, including a longer time sequence for angular momentum.  Note different time intervals.
    Initial conditions $\bm{P}=[0,0.0001,5.2]$, $\mathbf{L}=[0.0001,1,0.0001]$, with $\bm{P} \cdot \bm{P}=27.04$, $\bm{L} \cdot \bm{P}=0.0006$, $E=4.5327$, $\frac{\bm{P} \cdot \bm{P}}{EM_s}=5.9655$ (Zone 10). }
    \label{flutterchaotic}
\end{figure}

\begin{figure}[h]
    \centering
    \includegraphics[width=6in]{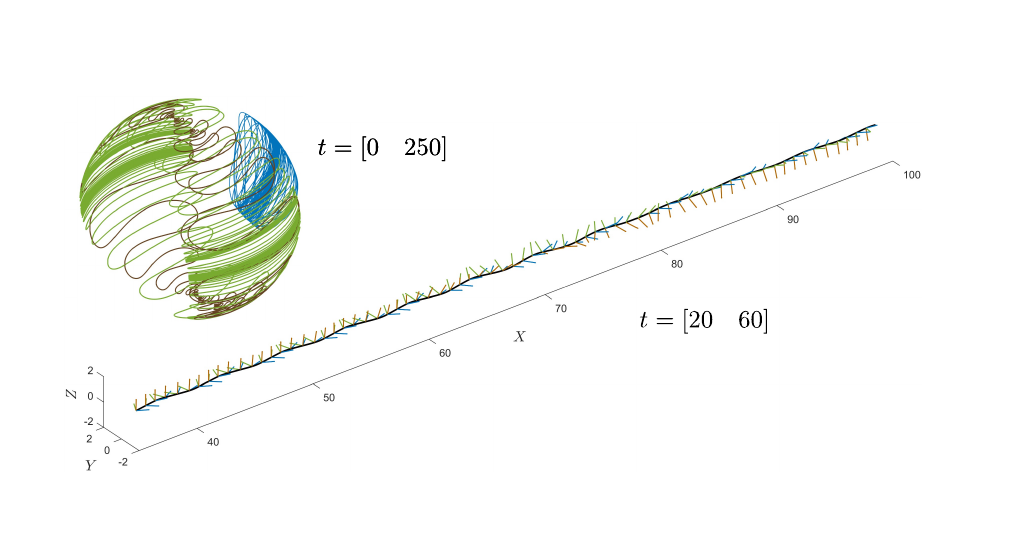}
    \vspace{-1cm}
    \caption{A single spin around an unstable $b$ axis fluttering mode $\text{F}_{b,c}$, with conditions as in Figure \ref{flutterregular}.      Evolution of the body axes, scaled according to body dimensions,  along a center of mass trajectory in real space, and of the body frame on a unit sphere. 
    Axes $a, b, c$ are in green, brown, blue. Note different time intervals.}
    \label{flutterregularspace}
\end{figure}

In Zone 10, there remain only two flutter modes, of which $\text{F}_{b,c}$ is unstable, to spinning associated with a change of sign of $L_b$, which is oscillating about zero. 
Plotting $P$ and $L$ separately gives this the appearance of a homoclinic connection, although it is not.  
 After making an excursion, the system returns to move on the same curves, but with different relative phase. 
Figure \ref{flutterregular}  is an example of regular spinning.
The fluttering mode $\text{F}_{b,c}$ corresponds to a line segment in $L$ space going through the origin, and an arc in $P$ space. 
The body periodically turns 
 in the same direction. 
Figure \ref{flutterchaotic} is an example of chaotic spinning, in which the body irregularly switches between branches in which the $L_c$ components have opposite signs. 
Both of these examples are far from the bifurcation between Zones 10 and 11, where the mode stabilizes. 
We note that when perturbations were chosen such that $\bm{L} \cdot \bm{P}$ remained strictly zero, only regular spinning was observed. 
The orientation of the body during a spin is shown in Figure \ref{flutterregularspace}. 
The $b$ axis takes a zigzag path from one pole to the other, and the $a$ axis from one arc to another, while the $c$ axis returns to the same arc, oscillating around the direction of motion. 
Again, the center of mass trajectory is three-dimensional. 

\clearpage

\section{Collections of trajectories with vanishing angular invariant}\label{fillspace}

In this section, we begin to look more systematically at the observed trajectories by seeing how they nest together to fill angular momentum space, and how saddles are connected in this space.  
We use the planar mode stability zones of Figure \ref{bifurcation} in Section \ref{planarmodestability} as a guide.  
Some of what we have already observed in the previous Section \ref{examplesolutions} suggest the presence of additional bifurcations in connectivity beyond the stability bifurcations separating these zones, as well as the presence of additional potentially simple, integrable landmark orbits in elliptic regions between saddles. 

Even a small nonzero angular invariant $\bm{L} \cdot \bm{P}$ tends to smear out saddle separatrices, and some chaotic multi-branch behavior was not observed when a planar system was specially perturbed so as to strictly preserve the $\bm{L} \cdot \bm{P}=0$ constraint. For these reasons, we examine the submanifold of solutions with this constraint. In each zone, we further hold linear momentum and energy constant when sampling across initial conditions. 

In Zone 1, only the intermediate axis tumbling modes $\text{T}_b$ are unstable. The system qualitatively resembles its limit with vanishing $\bm{P}$, the purely rotating rigid body, for which the trajectories are intersections of energy ellipsoids and angular momentum spheres in $L$ space alone.  
The magnitude of the angular momentum $\bm{L} \cdot \bm{L}$ is not conserved, but as $\bm{P}$ is relatively small, the system effectively wiggles around these trajectories.  Instead of two saddle points, the two $\text{T}_b$ modes are short line segments in $L$ space with saddle-like character. 

In Zone 2, both the intermediate and short axis tumbling modes $\text{T}_b$ and $\text{T}_c$ are unstable.  
As $\bm{P}$ is still relatively small, the trajectories effectively remain wiggling around near an ellipsoid in $L$ space, which is an approximation of the full ellipsoid in $P$-$L$ space. 
Now that there are two saddle pairs, four new elliptic-like regions exist in between, in the $L_b$-$L_c$ plane at  symmetrically oblique angles from the principal directions. 
As we approach near these ``centers'', we find solutions in which $L_a$ has small oscillations around zero, and $L_b$ and $L_c$ display two scales of oscillation, of small and medium amplitude,  
about finite values. 
The trajectories look like quasiperiodic orbits on a torus resembling a thickened, rounded cylinder. 
It appears possible to collapse these into points where pairs of $\pm L_b$ and $\pm L_c$ are equal by degenerating the ellipsoid into a spheroid. 

\begin{figure}[h]
    \centering
    \includegraphics[width=6in]{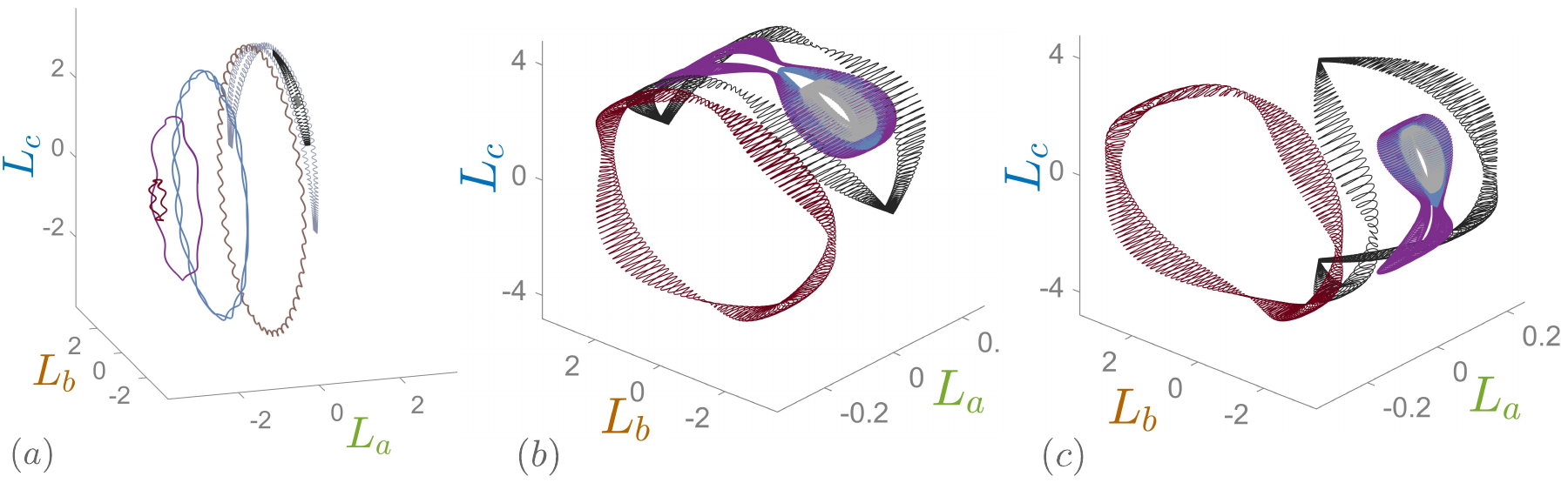}
    \caption{Trajectories in angular momentum space in Zone 2. (a) Overview. (b) and (c) Close up of saddle-center structure near the $L_b$-$L_c$ plane/equator, plotted with distorted aspect ratio, on either side of a connectivity bifurcation.  Details in text.  
    For all cases, $\bm{L} \cdot \bm{P}=0$ and $E=15$. 
    For both (a) and (b), $\bm{P} \cdot \bm{P}=34.5$ and $\frac{\bm{P} \cdot \bm{P}}{EM_s}=2.3$. For (c), $\bm{P} \cdot \bm{P}=35.1$ and $\frac{\bm{P} \cdot \bm{P}}{EM_s}=2.34$.\\
    Initial conditions 
for (a): 
reddish-brown $\bm{P}=[0,-5.8736,0]$, $\bm{L}=[-2.7678527,0,0.5]$,  
purple $\bm{P}=[0,-5.8736,0]$, $\bm{L}=[-2.300,0,2]$,  
silver-blue $\bm{P}=[0,-5.8736,0]$, $\bm{L}=[-1.4611,0,2]$, 
coffee-brown $\bm{P}=[0,-5.8736701,0]$, $\bm{L}=[-0.1951455,0,3.51]$, 
light grey $\bm{P}=[-5.8736,0,0]$, $\bm{L}=[0,-3.2901,0.01]$, 
black $\bm{P}=[-5.8736,0,0]$, $\bm{L}=[0,-0.01,3.38]$, 
dark grey $\bm{P}=[-5.8736,0,0]$, $\bm{L}=[0,-1.9872,2.7]$ ; \\
for (b): 
reddish-brown $\bm{P}=[0,-5.8736701,0]$, $\bm{L}=[-0.1951455,0,3.51]$, 
black $\bm{P}=[-5.8736,0,0]$, $\bm{L}=[0,-3.2901,0.01]$, 
purple $\bm{P}=[5.8736701,0,0]$, $\bm{L}=[0,-2.6973762,1.94]$, 
blue $\bm{P}=[-5.8736,0,0]$, $\bm{L}=[0,-0.01,3.38]$, 
grey $\bm{P}=[5.8736701,0,0]$, $\bm{L}=[0,-2.5021,2.2]$ ; \\
for (c): 
reddish-brown $\bm{P}=[0,-5.9245253,0]$, $\bm{L}=[-0.1986177,0,3.49]$, 
black $\bm{P}=[-11.2916,0,0]$, $\bm{L}=[0,1.0734,0.01]$, 
purple $\bm{P}=[-5.9245,0,0]$, $\bm{L}=[0,-2.3857,2.3]$, 
blue $\bm{P}=[-5.9245,0,0]$, $\bm{L}=[0,-3.2680,0.01]$,  
grey $\bm{P}=[-5.9245253,0,0]$, $\bm{L}=[0,-3.1204533,1]$.  
     }
    \label{zone2fol}
\end{figure}

Furthermore, there are two possible connectivities of the four saddles, separated by a bifurcation.
On one side of the bifurcation, surrounding the elliptic-like regions are heteroclinic $\text{T}_b$ flip orbits, either staying in one $c$ hemisphere or visiting both, homoclinic single-lobe $\text{T}_c$ orbits, 
and double-lobe orbits surrounding these and approaching close to $\text{T}_c$. 
On the other side of the bifurcation, the two axes switch roles, and there exist heteroclinic $\text{T}_c$ flip orbits, either staying in one $b$ hemisphere or visiting both, homoclinic single-lobe $\text{T}_b$ orbits, 
and double-lobe orbits approaching close to $\text{T}_b$. 
In between, close to the bifurcation, all of these orbits can come close to both axes.  For example, we see solutions that go close to a $\text{T}_c$ and two $\text{T}_b$s on one side, and a $\text{T}_b$ and two $\text{T}_c$s on the other.  Earlier, Figure \ref{tbhomo} was an example of a homoclinic solution near a transition. On either side, we also see solutions going close to four saddles, two each of $\text{T}_b$ and $\text{T}_c$. 
These behaviors are illustrated in Figure \ref{zone2fol} using several trajectories in $L$ space.  In Figure \ref{zone2fol}(a), a selection of orbits are seen to wiggle around an approximate energy ellipsoid, of which a bit more than one half is indirectly shown. In Figure \ref{zone2fol}(b)-(c), we zoom in and distort the aspect ratio near the saddle-center structure near the $L_b$-$L_c$ equator, to show the two possible connectivities. 

\begin{figure}[h]
    \centering
    \includegraphics[width=3in]{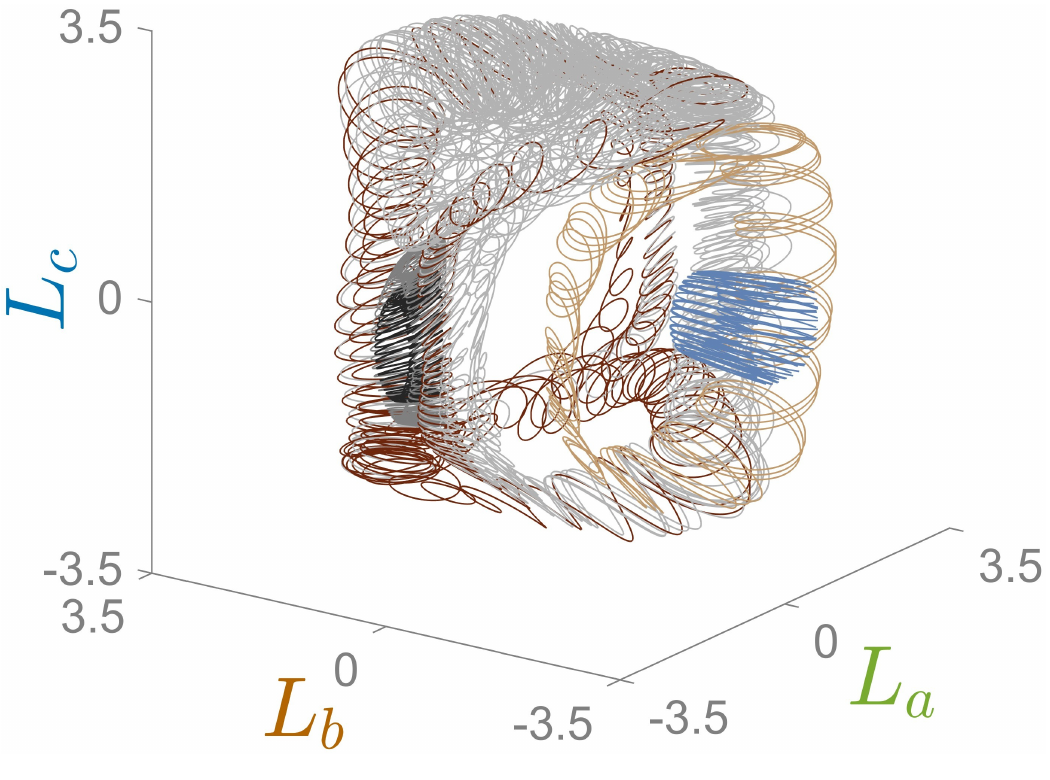}
    \caption{Trajectories in angular momentum space in Zone 5. Details in text. 
     For all cases, $\bm{P} \cdot \bm{P}=192$, $\bm{L} \cdot \bm{P}=0$, $E=40$, $\frac{\bm{P} \cdot \bm{P}}{EM_s}=4.8$. 
     Initial conditions: 
blue $\bm{P}=[-13.8564,0,0]$, $\bm{L}=[0,-2.1796,0.01]$, 
light brown $\bm{P}=[-8.1741,-11.1642,-0.7373]$, $\bm{L}=[1.74865, -1.36152,1.22945]$,
light grey $\bm{P}=[-13.8564,0,0]$, $\bm{L}=[0,-0.41804,2.2]$, 
reddish-brown $\bm{P}=[-13.8564,0,0]$, $\bm{L}=[0,-0.9833,2]$, 
dark grey $\bm{P}=[0,-13.8564,0]$, $\bm{L}=[0,1.0734,0.01]$, 
black $\bm{P}=[0,-13.8564,0]$, $\bm{L}=[-2.3307,0,1.2]$. 
}
    \label{zone5fol}
\end{figure}

In Zone 3, only the short axis tumbling modes $\text{T}_c$ are unstable.  Example trajectories were shown earlier in Figures \ref{tcflipreg}-\ref{tcflipchao}.  The approximate energy ellipsoid is still visible. 
In Zone 4, all three tumbling modes are stable. 
In Zone 5, only the short axis tumbling modes $\text{T}_c$ are unstable.
Figure \ref{zone5fol} shows several angular momentum trajectories, including regular motion near the elliptic-like axes, chaotic ``hair'' around the unstable axis, and travel through a rectilinear ``cage'' of tunnels in between.  On the $L_a$-$L_c$ plane, the elliptic-like regions around the $a$-axis tumbling modes are slightly re-entrant, meeting up with the unstable modes in cusp-like features.  
These perhaps collide in the transition to Zone 6, where both the long and short axis tumbling modes $\text{T}_a$ and $\text{T}_c$ are unstable. 

\begin{figure}[h]
    \centering
    \includegraphics[width=6in]{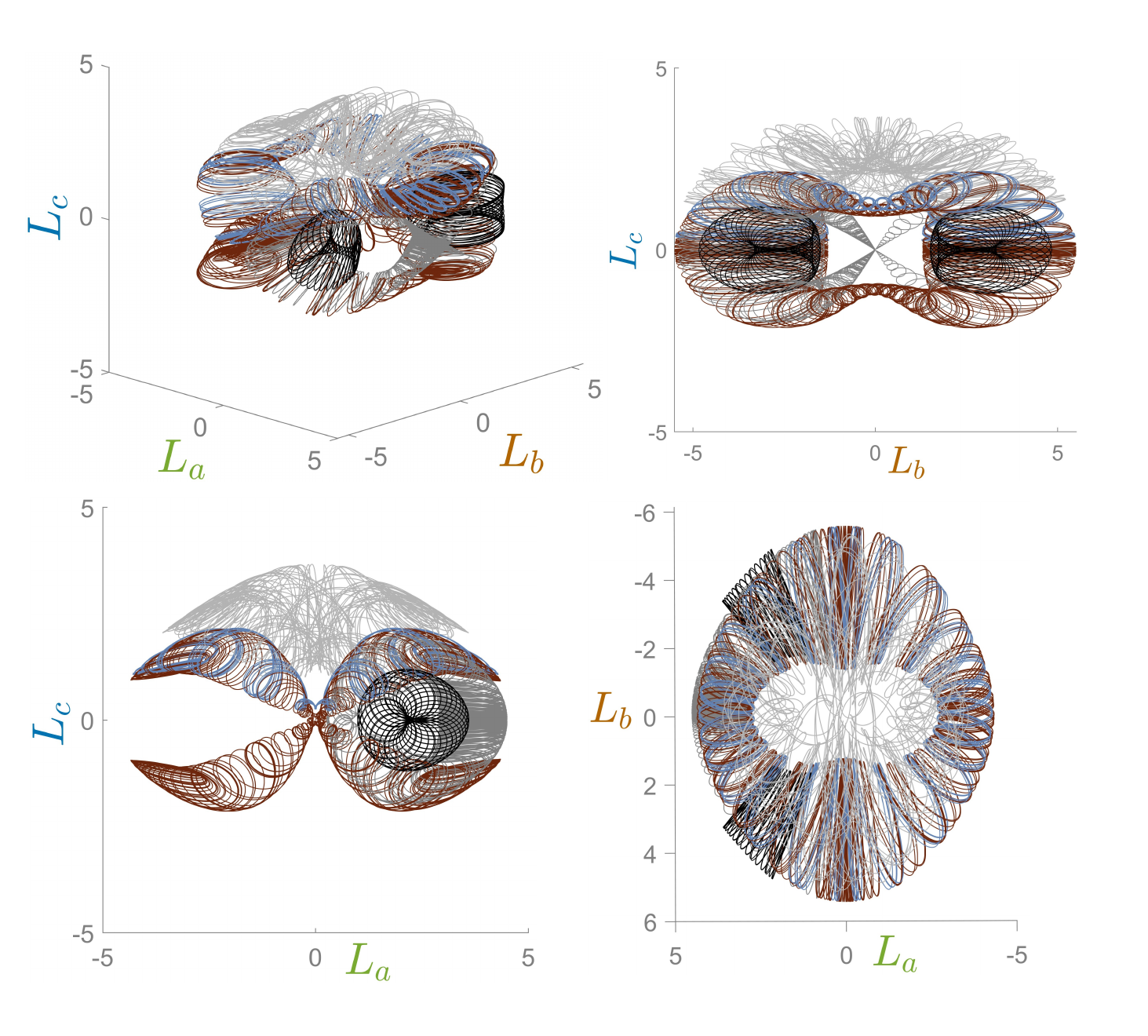}
    \vspace{-0.5cm}
    \caption{Trajectories in angular momentum space in Zone 7, in oblique view and projected onto the three principal planes.   Details in text.  For all cases, $\bm{P} \cdot \bm{P}=446.3902$, $\bm{L} \cdot \bm{P}=0$, $E=85.6795$, $\frac{\bm{P} \cdot \bm{P}}{EM_s}=5.21$. 
    Initial conditions: 
black (two) $\bm{P}=[\mp 11.4025,-11.4025,-13.6511]$, $\bm{L}=[-2.6,\pm 2.6,0]$, 
dark grey $\bm{P}=[0,-21.1279,0]$, $\bm{L}=[2.9133,0,0.02]$, 
reddish-brown $\bm{P}=[-21.1279,0,0]$, $\bm{L}=[0,-1.2942,0.01]$, 
silver-blue $\bm{P}=[-21.1279,0,0]$, $\bm{L}=[0,1.2713260,0.25]$, 
light grey $\bm{P}=[-21.1279,0,0]$, $\bm{L}=[0,0.2854,1.3]$. 
}
    \label{zone7fol} 
\end{figure}

In Zone 7, all three tumbling modes are unstable. 
Trajectories around the short $c$ axis remain complex, resembling hair, while those around the long and intermediate $a$ and $b$ axes have a clearer saddle-like structure. Four new elliptic-like regions exist in between these, in the $L_a$-$L_b$ plane at symmetrically oblique angles from the principal directions. 
As we approach near these ``centers'', we find solutions in which $L_c$ has large two-scale oscillations around zero, and $L_a$ and $L_b$ oscillate with non-constant amplitude 
about finite values. 
Figure \ref{zone7fol} shows several angular momentum trajectories, in oblique view and projected onto the three principal planes. 

This zone also undergoes a bifurcation in connectivity between the four saddles, analogous to that in Zone 2. 
On one side of the bifurcation, as in Figure \ref{zone7fol}, surrounding the elliptic-like regions or the ``hair'' are $\text{T}_a$ flip orbits (either two branches via $b$, two branches via $c$, or four branches via both), and figure-eight trajectories around $\text{T}_b$, of which Figure \ref{doublebhomo} was an earlier example. 
On the other side of the bifurcation, the two axes switch roles, and there exist $\text{T}_b$ flip orbits (either two branches via $a$, two branches via $c$, or four branches via both), and figure-eight trajectories around $\text{T}_a$. 
In between, close to the bifurcation, we see solutions approaching near three or four $a$ or $b$ saddles in one or two hemispheres, as in the examples in Figures \ref{tbpmta}, \ref{tapmtb}, and \ref{pmtapmtb2}. 

\begin{figure}[h]
    \centering
    \includegraphics[width=6in]{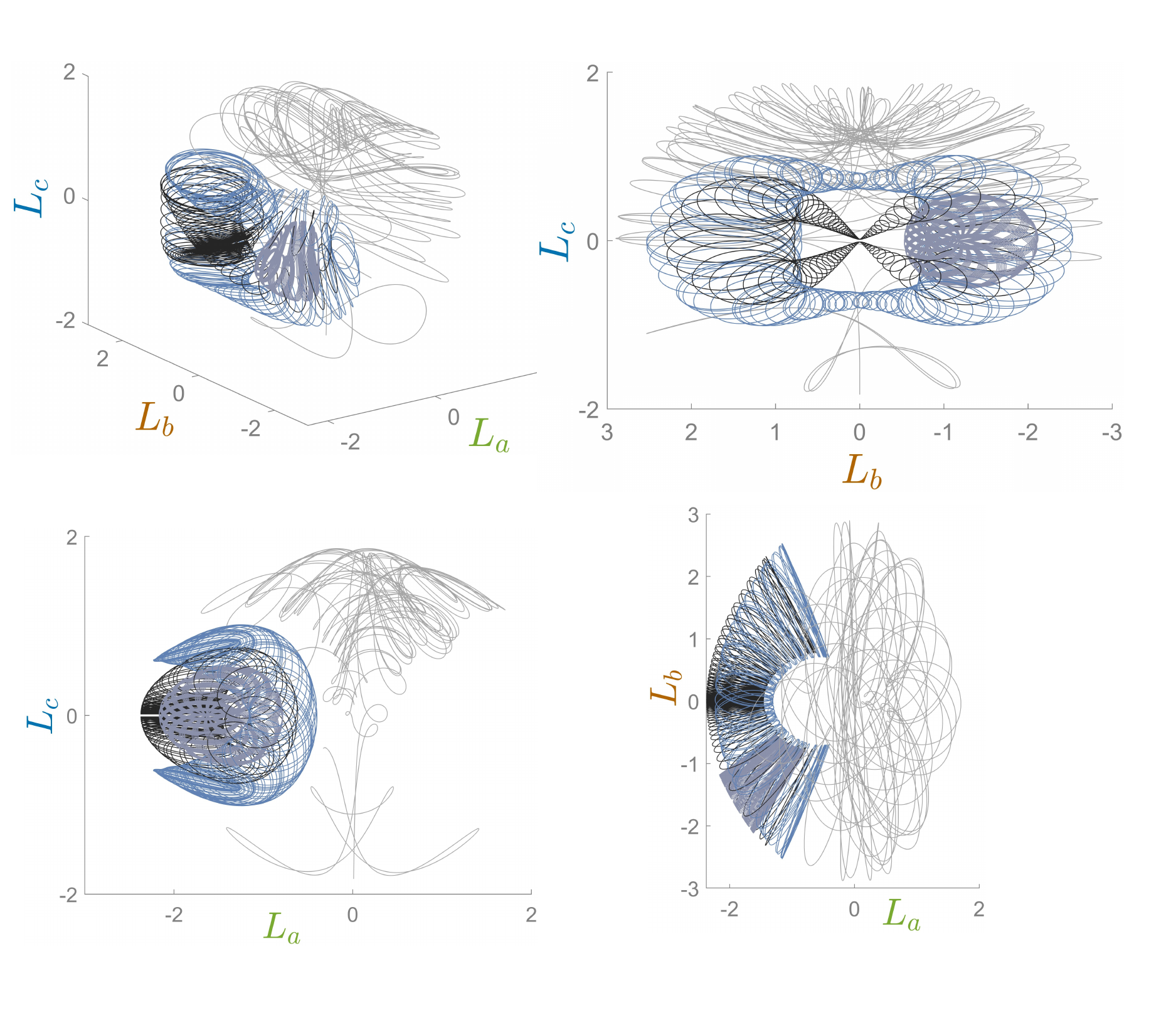}
    \vspace{-1cm}
    \caption{Trajectories in angular momentum space in Zone 8, in oblique view and projected onto the three principal planes.   Details in text.  For all cases, $\bm{P} \cdot \bm{P}=132.75$, $\bm{L} \cdot \bm{P}=0$, $E=25$, $\frac{\bm{P} \cdot \bm{P}}{EM_s}=5.31$. 
    Initial conditions:
dark grey $\bm{P}=[4.6571,-7.4513,-7.4523]$, $\bm{L}=[-1.6,-1,0]$,  
black $\bm{P}=[0,-11.5217,0]$, $\bm{L}=[-1.4538,0,0.02]$,
silver-blue $\bm{P}=[0,-11.5217,0]$, $\bm{L}=[-1.13075,0,0.8]$, 
light grey $\bm{P}=[0,-11.5217,0]$, $\bm{L}=[0.01,0,-1.82948]$.  
}
    \label{zone8fol}
\end{figure}

In Zone 8, the long axis tumbling mode $\text{T}_a$ and the intermediate and short axis flutter modes $\text{F}_{b,c}$ and $\text{F}_{c,b}$ are unstable. 
 Figure \ref{zone8fol} shows several angular momentum trajectories, in oblique view and projected onto the three principal planes. 
These include figure-eight trajectories around $\text{T}_a$, a new ``center'' in the $L_a$-$L_b$ plane, and a chaotic spinning trajectory around $\text{F}_{b,c}$. 

\begin{figure}[h]
    \centering
    \includegraphics[width=6in]{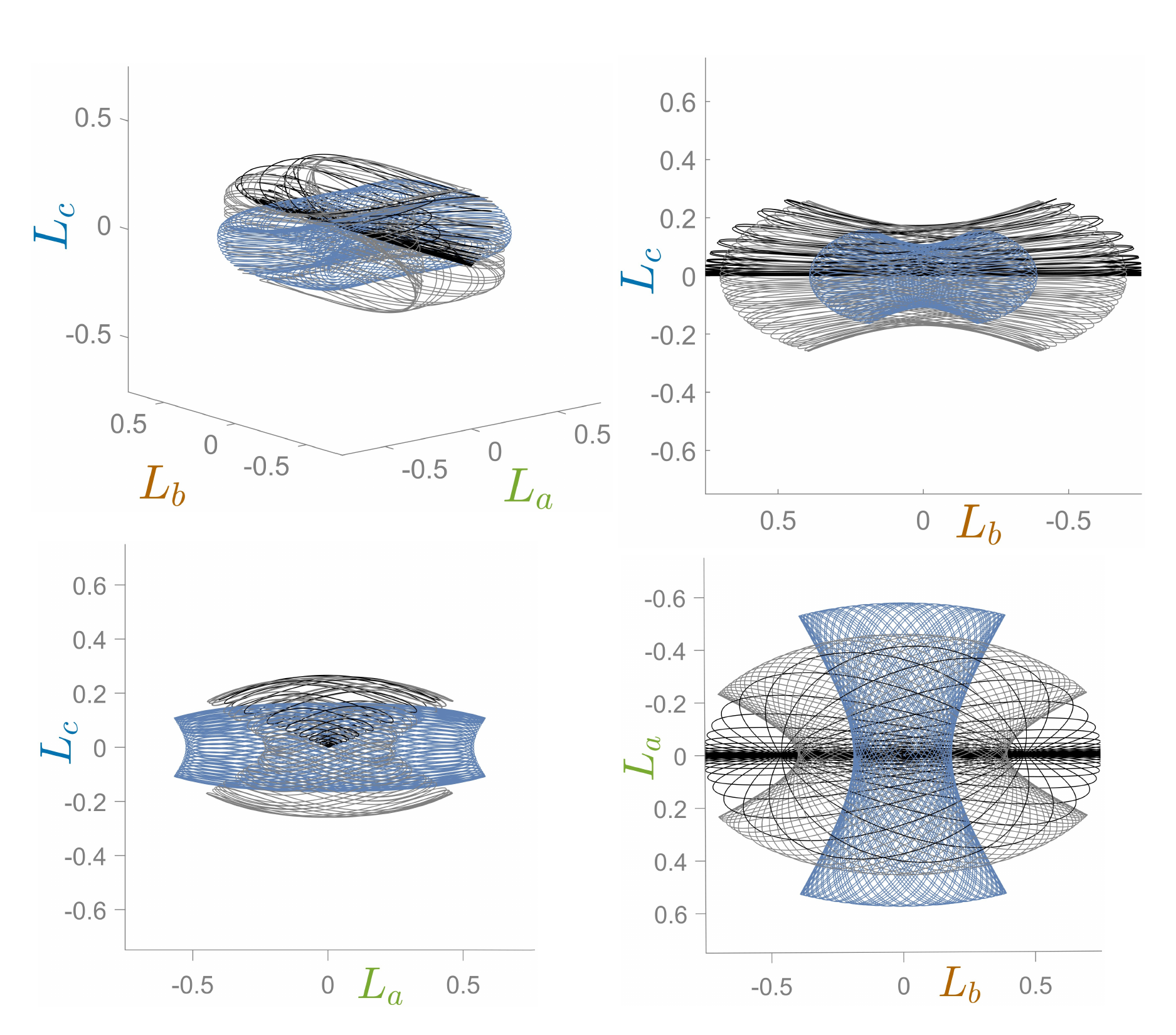}
    \vspace{-0.5cm}
    \caption{Trajectories in angular momentum space in Zone 10, in oblique view and projected onto the three principal planes.   Details in text.  For all cases, $\bm{P} \cdot \bm{P}=29.4625$, $\bm{L} \cdot \bm{P}=0$, $E=4.5327$, $\frac{\bm{P} \cdot \bm{P}}{EM_s}=6.5$. 
    Initial conditions:
silver-blue $\bm{P}=[0,1.5564,5.2]$, $\bm{L}=[-0.4869,-0.3340,0.1]$,  
grey $\bm{P}=[1.5565, 0, 5.1999]$, $\bm{L}=[-0.3341,0.4892,0.1]$, 
black $\bm{P}=[2.1124,0,5]$, $\bm{L}=[-0.0023,-0.5332,0.001]$
    }
    \label{zone10fol}
\end{figure}

In Zone 9, the long axis tumbling mode $\text{T}_a$ is stable, its elliptic-like region coexisting with spinning trajectories 
around the unstable intermediate and short axis flutter modes $\text{F}_{b,c}$ and $\text{F}_{c,b}$. 
In Zone 10, the long axis flutter mode $\text{F}_{a,c}$ is stable, the intermediate axis flutter mode $\text{F}_{b,c}$ is unstable, and the short axis flutter mode no longer exists. 
Figure \ref{zone10fol} shows several angular momentum trajectories for stable and unstable flutter, in oblique view and projected onto the three principal planes. 
Figures \ref{flutterregular} and \ref{flutterchaotic} were earlier examples of unstable flutter in this zone. 
In Zone 11, these two remaining flutter modes are stable. 

\clearpage

\section{Discussion}\label{discussion}

The trajectories in Sections \ref{examplesolutions} and \ref{fillspace} reveal aspects of the structure of the solution manifold for the Kirchhoff equations for an Aref-Jones ellipsoid, particularly for the $\bm{L} \cdot \bm{P}=0$ submanifold. While the system is chaotic, much of the chaos manifests in irregular sequences of paths between well-defined saddle-like structures.  
Still, much of the space remains unexplored. In between the pure and planar landmarks, there is an ocean of unknown behaviors with intermediate values of $\bm{L} \cdot \bm{P}$.  The $\bm{L} \cdot \bm{P}=0$ submanifold is not integrable in general, and we observed loss of integrability for some of its regular solutions when moving even slightly off of this submanifold.  

Arranging the solutions as in Section \ref{fillspace} indicates what appear to be new integrable solutions at the oblique junctions between the regions surrounding the planar tumbling states. 
In Zones 2 and 7 these new solutions are marginally stable, sitting between unstable planar states. 
Degenerating the body properties onto those of a spheroid appears to allow for analytical solutions of comparable simplicity to the planar solutions, but we did not explore this issue in depth or attempt comparison with special solutions in the literature on symmetric bodies. 
In, for example, Zone 5, there is the suggestion of new unstable states in between stable planar states. 
In this zone, there is no obvious saddle structure in the $L_a$-$L_b$ plane, but there is an interesting cuspy saddle-like structure in the $L_b$-$L_c$ plane. 

The seeming simplicity of Zone 4, which we did not explore as all tumbling states were stable, might actually belie something more interesting, as it is not obvious how these stable orbits meet up to fill the space. 
This zone is interesting in and of itself. There is an intermediate ratio of linear to angular momentum for which planar tumbling around any axis is stable.  This should have implications for launching projectiles, particularly if this structure persists with the addition of gravity. 

Zones 2 and 7 exhibit bifurcations in connectivity.  It should be possible to determine their location in terms of the linear momentum to energy ratio, as they require that planar tumbling solutions around different axes have the same energy.  These and other connectivity features in the case of vanishing angular invariant are easily blurred by even a small amount of this quantity. 
It is possible that other such bifurcations exist in other zones, but are not easy to see, and such questions would benefit from a more systematic mathematical approach to the topology of the intersections of conserved quantity level sets.

In some zones, we observed unstable tumbling trajectories confined within figure-eight or more complex shaped regions, where the body returns to the same location near an unstable solution in $L$ space after taking two or more different paths.  We did not explore these behaviors in detail, but it is likely that the body is not returning to the same state in the full six-dimensional space.  
Two issues are related. Our representation using separated linear $P$ and angular $L$ momentum spaces is a choice that obscures some aspects of the dynamics.  In particular, hyperbolic $\bm{L} \cdot \bm{P}$ type isosurfaces are not visible, and we are not able to clearly distinguish homo- and heteroclinic connections when the distinction between them is the relative phase of motion on curves in $P$ and $L$ spaces.  In an example of flutter unstable to spinning, we observed a phase shift corresponding to a $\pi$-flip of the third (unlettered) axis of the body.  Because of the symmetry of the body in the present case, this could be thought of as the same state, but this would mean that we identify points in $\mathbb{R}^6$ so that the system moves in a differently connected space.  
These issues remain to be clarified and further explored, in part perhaps by judicious use of Poincar{\'{e}} sections. 

Given that the approximate energy ellipsoid is visible in angular momentum $L$ space when the linear momentum $P$ is low, perhaps it would also be possible to see this structure emerging in linear momentum space when $P$ is dominant. 

We did not explore the center of mass spatial trajectories of the bodies, instead focusing on the behavior of the components and the body frame, although the latter is related. 
 While these trajectories can be complex three-dimensional motions, the amplitudes observed were very small and would have required distortion of the aspect ratio of trajectories in order to see them clearly. 
It would be interesting to explore possible connections between spinning away from unstable flutter and the regular precession under gravity seen in \cite{Auguste13}, or to relate the larger amplitude helical motions seen in \cite{Zhong11, Varshney13} to the present study.  
Trajectories become of greater interest and importance when there is a direction singled out, by gravity for example.  Recall that for the tumbling and fluttering solutions of Lamb, the former are at an angle with respect to gravity whereas the latter follow overall the direction of gravity. 

Another potentially fruitful area of comparison with our fully three-dimensional problem is the planar problem with gravity \cite{Kozlov89, Borisov06} and dissipation \cite{Kuznetsov15} or additional features \cite{PomerenkRistroph25}. 
Recently, Pomerenk and Ristroph \cite{PomerenkRistroph25} mapped out a variety of spatial trajectories of a reduced model, more complex than Kirchhoff, for planar motions of a plate with noncoincident centers of mass and buoyancy.  
It would likely be illustrative to explore relations between planar stability of simple motions such as ``diving'', ``pancaking'', and ``gliding'', and three-dimensional stability of pure and fluttering modes, particularly how stability depends on body shape and mass inhomogeneity, and how these states are connected in the six-dimensional space. 

Our study has been restricted to a particular, previously studied, simple shape.  More complex shapes likely admit additional integrable solutions, and live in different spaces because they have different energy isosurfaces--- for example, a propeller tensor adds cross terms in the expression for the energy. Additional ``mixed-mode'' solutions, and differences in pure mode stability and connectivity, were shown for ellipsoids with non-standard ordering in \cite{Holmes98}. 
A systematic study of the explicit relations between body type, including fore-aft asymmetry, nonconvexity, and chirality, and the various tensors and buoyancy effects would be useful. 

It may be worth exploring an analogy between this problem and the dissipative dynamics of small bodies in viscous flows, for which body-dependent resistance tensors relate velocities and forces, and body and flow symmetries lead to conserved quantities. The most well-known integrable motions in this system are periodic tumbling of ellipsoids in shear flows, called  Jeffery orbits \cite{Jeffery22}, but several isolated examples have been found for the more general problem \cite{Ishimoto23}. In an early paper exploring the relations between body type and resistance tensors, Brenner \cite[Section 8]{Brenner64pt2} suggested an analogy between the Kirchhoff kinetic energy and the Jeffery dissipation.  

\section{Conclusions}

We have observed aspects of the behavior of a rigid body in an ideal fluid, by examining the solutions of the Kirchhoff equations for a simple ellipsoidal body. The stability of known integrable solutions, including tumbling and fluttering, was determined and used as a guide for further exploration. A variety of behaviors were documented, including flipping and spinning between unstable states, and chaotic sequences of paths between such states. We investigated how trajectories with vanishing angular invariant fill momentum space, and indirectly showed how saddles and regions of this space are connected. This process uncovered additional connectivity bifurcations and apparently new integrable solutions. Many questions remain, and the space of solutions with general values of the angular invariant remains mostly unexplored.

\section*{Acknowledgments}

We thank A. Dehadrai for help, S. Jabuka and O. Pomerenk for discussions, and M. Aureli for help, discussions, and tolerating, to the point of invitation, repeated incursions into his office to borrow his copy of Lamb.

\bibliographystyle{unsrt} 

 \end{document}